# Extreme antagonism arising from gene-environment interactions

Thomas P. Wytock[1], Manjing Zhang[2], Adrian Jinich[3], Aretha Fiebig[4], Sean Crosson[4] and Adilson E. Motter[1,5,6,*]

[1]Department of Physics and Astronomy, Northwestern University, Evanston, IL 60208, USA
[2]The Committee on Microbiology, University of Chicago, Chicago, IL 60637, USA
[3]Division of Infectious Diseases, Weill Department of Medicine, Weill−Cornell Medical College, New York, NY 10065, USA
[4]Department of Microbiology and Molecular Genetics, Michigan State University, East Lansing, MI 48823, USA
[5]Chemistry of Life Processes Institute, Northwestern University, Evanston, IL 60208, USA
[6]Northwestern Institute on Complex Systems, Northwestern University, Evanston, IL 60208, USA
* To whom correspondence should be addressed: motter@northwestern.edu

## Abstract
Antagonistic interactions in biological systems, which occur when one perturbation blunts the effect of another, are typically interpreted as evidence that the two perturbations impact the same cellular pathway or function. Yet, this interpretation ignores extreme antagonistic interactions wherein an otherwise deleterious perturbation compensates for the function lost due to a prior perturbation. Here, we report on gene-environment interactions involving genetic mutations that are deleterious in a permissive environment but beneficial in a specific environment that restricts growth. These extreme antagonistic interactions constitute gene-environment analogs of synthetic rescues previously observed for gene-gene interactions. Our approach uses two independent adaptive evolution steps to address the lack of experimental methods to systematically identify such extreme interactions. We apply the approach to *Escherichia coli* by successively adapting it to defined glucose media without and with the antibiotic rifampicin. The approach identified multiple mutations that are beneficial in the presence of rifampicin and deleterious in its absence. The analysis of transcription shows that the antagonistic adaptive mutations repress a stringent response-like transcriptional program, while non-antagonistic mutations have an opposite transcriptional profile. Our approach represents a step toward the systematic characterization of extreme antagonistic gene-drug interactions, which can be used to identify targets to select against antibiotic resistance.

## Statement of Significance
Mutations that are deleterious in the absence of an antibiotic can become beneficial in its presence, which is an example of an extreme antagonistic gene-environment interaction. Such antagonism is of biophysical significance because it reflects non-local interactions mediated by intracellular networks rather than direct physical or chemical interactions. We develop and apply a forward-evolution experimental approach to systematically identify these interactions using DNA and RNA sequencing. These analyses reveal differences in the expression of the translational machinery between antagonistic and non-antagonistic mutations. Our findings demonstrate how distinct single-base pair mutations within the same gene can have divergent phenotypic consequences, which give rise to antagonistic interactions that can be explored in addressing antibiotic resistance.

# Introduction

Complex biological functions generally emerge from interactions rather than solely from the contributions of individual genes and environmental factors *(1)*. When the function of interest is growth rate, as often considered in fitness studies of single-cell organisms and immortalized cell lines, interactions are typically characterized in terms of whether they enhance or suppress fitness beyond what is predicted from the individual contributions. In the case of gene-gene and drug-drug interactions, pairwise relationships have been systematically identified using gene knockouts and other function-impairing perturbations that effectively query combinatorial effects, through large-scale genetic screens *(2-6)*, metabolic modeling *(7-9)*, and collateral sensitivity assays *(10-12)*. Interactions are classified as synergistic if the double-perturbation fitness is lower than expected from adding the effects of the individual perturbations and antagonistic if higher *(13)*. Synergistic and antagonistic interactions are often interpreted as the result of local mechanisms *(2, 14, 15)*. In this interpretation, synergistic interactions would relate genes in parallel or redundant pathways. Antagonistic interactions, on the other hand, would concern serial pathways of non-redundant genes or drug targets contributing to the same process. In the latter case, since the loss of either gene alone hampers the process, antagonism would emerge because the fitness impact of the second knockout is directly limited by the first.

An interesting exception to this local picture is the case of synthetic rescues, which are extreme antagonistic interactions in which a gene knockout becomes beneficial (rather than merely less deleterious) when applied after the knockout of another gene *(16-19)*. Synthetic rescues can be interpreted as a form of sign epistasis *(20)* between two genetic perturbations that are each individually deleterious, in the sense that the fitness impact of one perturbation changes sign and becomes beneficial in the presence of the other. While some synthetic rescues can be attributed to toxicity *(21-26)*, previous work indicates that there exists a combinatorial number of synthetic rescues that are mediated by biochemical networks and involve genes in disparate processes *(17, 27)*. Even though synthetic circuits have been used to study the evolution of extreme antagonism *(28, 29)*, and the identification of gene-gene combinations exhibiting non-local synthetic rescues is ongoing *(18, 19)*, an open proposition is whether analogous extreme antagonistic interactions can be systematically identified for gene-environment combinations. Determining the prevalence of extreme antagonism would address a fundamental question in biology concerning the nature of possible interactions between genes and environmental factors. Furthermore, implementing such environmental factors using antibiotic stressors would offer a pathway to design new antibiotic combinations that exhibit collateral sensitivity, as discussed below, because the action of one antibiotic disfavors the acquisition of resistance to another *(30, 11, 31-36)*.

Here, we explore a method to systematically identify extreme antagonistic gene-environment interactions. These interactions occur between genetic and environmental perturbations that, individually, decrease fitness relative to the original condition but that, when combined, increase fitness relative to the environmental change alone. In parallel with synthetic rescues vis-à-vis sign epistasis, extreme antagonistic gene-environment interactions are an outstanding class of environmental sign epistasis *(37)*, in that mutations that are deleterious in a permissive environment become beneficial in a more restrictive environment. We identified these interactions using serial adaptive evolution of *Escherichia coli*, which consisted of 1) adaptation to a defined glucose medium yielding strains with enhanced fitness, and 2) subsequent adaptation of these strains to antibiotic stress defined by a sublethal concentration of rifampicin (rif). The gene-environment interactions were characterized by comparing the growth fitness of all strains grown in both the presence and the absence of rif, quantifying the cost of resistance *(28)*. Previous studies of the cost of rif resistance have focused on strains cultivated in complex media and reported mutations in the beta subunit of RNA polymerase (*rpoB*) that result in a fitness disadvantage in such media in the absence of rif *(38-40)*, though adaptive lab evolution experiments in defined glucose media have proved the same mutations to be beneficial *(19)*. Because previously identified adaptive mutations that enhance growth rate and/or rif resistance have pleiotropic effects *(41-43)*, the global transcriptional consequences of adaptation were further investigated using RNA-sequencing (RNA-Seq) before and after each adaptive step. Our experiments and analyses identified mutations in rif-adapted strains that confer growth faster, equal, and slower in the absence of rif compared to that of the parent strain. Of these three groups, the mutations conferring slower growth exhibit extreme antagonistic interactions with rif. These differences in growth can be attributed to widespread transcriptional reprogramming, which we map to broader cellular processes including central metabolism, translation, and a stringent-like response.

We propose that extreme antagonistic gene-environment pairs in which the environmental perturbation is the addition of a drug, such as rif in our experiments, provide targets for the design of antibiotic combinations that can select against resistance. For a pair of extreme antagonistically interacting antibiotics, growth inhibition by a drug combination is weaker than that of one of the drugs alone, meaning that acquisition of resistance to a second drug would cause growth to be suppressed more strongly *(44, 45)*—a phenomenon referred to as collateral sensitivity *(11)*. While previous studies of collateral sensitivity have found some extreme antagonistic drug pairs *(30, 46, 32, 47, 43, 12)*, such pairs tend to be rare *(34)*, and are found by directly testing existing antibiotics.

In this paper, we apply our method to study antagonism as it relates to rif, for which there appear to be no antagonistic partner antibiotics *(11)*. Resistance to rif, as with antibiotic resistance in general, induces pleiotropic effects on cell physiology that mitigate the action of the drug but are otherwise suboptimal *(41, 48, 43),* which we are able to characterize transcriptionally using RNA-Seq in the presence of the drug (to discern adaptive changes) and in the absence of the drug (to identify costs of resistance). We expect transcriptional profiles to be predictive of collateral sensitivity, as has been shown for chemogenomic profiles *(49),* to the extent that both reflect the molecular mechanisms underlying adaptation and the cost of resistance. These efforts may offer another avenue to manage resistance in bacterial infections that are treated with rif, such as *Mycoplasma tuberculosis (50, 51)*. Our approach to find extreme antagonistic gene-drug pairs is scalable and can facilitate the systematic design of antagonistically interacting drug combinations by providing targets—the gene mutations—against which a second drug may be developed. Ultimately, our results provide insights into fundamental and applied aspects of gene-environment interactions.

## Materials and Methods
### Strain cultivation
All strains used in this study were derived from wild type (WT) *E. coli* strain K12 substrain BW25113 *(52)*. The WT strain was suspended in media supplemented with 0.4% w/v glucose (M9G) in triplicate and allowed to grow at 37 ºC with shaking for 12 hours, at which point cultures were diluted to an initial optical density at 600nm (OD$_{600}$) of 0.01. Growth and dilution proceeded in this manner for 21–28 days (216–611 generations) in an effort to identify cultures with an enhanced growth rate. At the end of this period of evolution, cultures were plated, colonies were recovered, and whole-genome sequencing (WGS) was performed, yielding Ref mutants with the mapped mutations *pykF*(C8Y) and *rpoB*(T1037P).

### Selection for spontaneous mutants that restore fast growth in the presence of rif

Mutations that enabled fast growth of Ref strains in M9G with 7.5 µg/mL rif (M9G + rif)—an inhibitory concentration—were selected according to the following procedure. Three independent 5 mL cultures of each of these three strains were cultivated in 20 x 150 mm borosilicate tubes and serially passaged with a starting $OD_{600}$ of 0.01 for 10 days, shaking at 37 °C inclined at a 45° angle at 200 rpm in an Infors Shaker. As before, strains were diluted every 12 hours to ensure the cells remained in exponential phase. Spontaneous acquisition of fast growth (i.e., rif resistance) mutations is random based on published data *(53)*. We observed an enhanced growth rate of the rif-containing cultures, as expected, within 4-5 days (8-10 passages; ≈60-70 generations). Enhanced growth rate of the culture was typically evident after 4-5 days. On the final day of serial passaging, these fast-growing cultures were streaked onto LB agar to isolate single clones. Growth rate measurements of clones confirmed that these "Mut" strains grew faster in M9G + rif than the Ref strains from which they were derived. The genetic identity of the Mut strains was verified by WGS.

### Growth rate measurements

To measure growth rates, overnight cultures were first started from freshly grown colonies in 2 mL M9G shaken at 37 °C overnight. These starter cultures were diluted in triplicate to $OD_{600}$ = 0.01 in 2 mL of M9G either with or without 7.5 µg/mL rif in 13 x 100 mm borosilicate tubes and were allowed to grow at 37 °C inclined at a 45° angle at 200 rpm in an Infors Shaker. The $OD_{600}$ was measured approximately every 20 minutes for at least 4 independent cultures. Cultures were grown in the absence of light to prevent it from inactivating rif. To obtain growth rates, the log-linear region of the growth curves was fit to the exponential growth equation $OD_{600}(t) = 2^{t/D}$, where *t* is time and *D* is the doubling time, measured in minutes. The mean and standard deviation were calculated for the doubling time ratio of each derived mutant to its corresponding Ref parent. Differences in growth were assessed using a *t*-test on the ratio of Mut strain growth to Ref strain growth to control for possible differences in media preparation (Table S1 in the Supporting Material).

### Mapping rif-adaptive mutations by WGS

We isolated genomic DNA from all *E. coli* mutant strains presented herein, using a standard guanidinium thiocyanate extraction and isopropanol/ethanol precipitation. Briefly, the DNA was randomly sheared and libraries were prepared for WGS using an Illumina HighSeq 4000 (50-bp single end reads). The WGS data from each strain was assembled to the *E. coli* BW25113 WT genome template, and polymorphisms were identified using the *breseq* analysis pipeline *(54)*. Each sequenced genome library yielded an average of 26 million reads, resulting in average depth of coverage greater than 250x.

### Measuring transcript levels by RNA-Seq

We isolated RNA for sequencing from both Ref strains and their rif-adapted Mut strains in triplicate, in both M9G and M9G + rif; cells were grown and harvested across three separate days.

The growth protocol to prepare cultures for RNA isolation was as follows. All strains were cultivated in M9G at 37 °C; starter cultures were diluted to $OD_{600} = 0.01$ with or without 7.5 μg/mL rif to an $OD_{600}$ of ~0.18-0.22. At that point, RNA was extracted by pelleting cells for 30 seconds and rapidly resuspending the pellets in 1 mL Trizol (Invitrogen, Life Technologies). RNA was extracted from the Trizol suspension using the manufacturer's protocol. The extracted RNA was next treated with Turbo DNase (Ambion, Life Technologies) and further purified using an RNA purification kit (Qiagen). Absence of DNA contamination was confirmed by PCR, where the lack of PCR product (about 100 bp in length) relative to a DNA-containing positive control was interpreted as evidence of DNA removal. RNA-Seq libraries were prepared with an Illumina TruSeq stranded RNA kit according to manufacturer's instructions. Sequencing (50 bp single-end read) was performed on an Illumina HiSeq 4000. Transcript levels were mapped to the *E. coli* BW25113 WT genome in CLC Genomics Workbench 11 (mismatch cost = 2; insertion cost = 3, deletion cost = 3, length fraction = 0.8, similarity fraction = 0.8). Sequencing reads for all strains, plus and minus rif treatment, have been deposited in the NCBI GEO database (accession number GSE136977).

## Annotation analysis of transcriptional changes

### Determination of differential expression.
We compared the transcript levels measured Mut strains with their respective Ref parents when grown in M9G or M9G + rif. Differential expression was assessed using DESeq2 with the "apeglm" shrinkage estimator yielding $\log_2$ fold changes and single-gene *p*-values *(55)*. These results were further interpreted in the context of Protein Analysis Through Evolutionary Relationships (PANTHER) pathway annotations, PANTHER protein classes, Clusters of Orthologous Groups (COG) identifiers, COG groups, modulons derived from independent component analysis of gene expression, and origons in the genetic network of the Regulon Database (RegulonDB).

### PANTHER analysis.
The $\log_2$ fold changes were input into PANTHER version 14.1 using the *E. coli* annotations for PANTHER pathways and PANTHER protein classes using a 5% FDR threshold for significance *(56)*.

### COG analysis.
The COG annotation scheme groups genes into 5 classes, which are subdivided into 23 categories comprising 2,161 identifiers based on protein amino acid sequence *(57)*. We interrogated the transcriptional $\log_2$ fold changes for enrichment of COG categories and IDs using a bootstrapping approach. For all *n* genes associated with a term, we calculated the rank sum and compared this with the rank sums derived from *N*=20,000 randomly selected sets of *n* genes. We calculated the number *X* of randomly generated rank sums that were smaller than the observed rank sum among the distribution of randomly generated rank sums and determined a *p*-value using $1 - |1 - 2X/N|$. After determining the *p*-values, we selected terms for further consideration using the Benjamini-Hochberg FDR procedure with a 5% threshold *(58)*.

### RegulonDB transcription factor analysis.
The network of 212 transcription factors and 1,814 regulated genes was defined in the "generegulation_tmp.txt" file downloaded from RegulonDB *(59)*. Annotation enrichment proceeded as before, with the caveat that fold changes were multiplied by the valence of the transcription factor-gene interaction before calculating the rank sum. In addition, any transcription factors passing the 5% FDR threshold but having $\log_2$ fold changes in conflict with their downstream targets were excluded from consideration. We record all significant terms and their signs in Table S2-S6 of the Supporting Material for all Mut strains in both M9G and M9G + rif.

### Quantifying the transcriptional signature of the stringent response

We re-analyzed the data contrasting the transcriptional outcomes of (stringent response-inducing) serine hydroxamate treatment between WT *E. coli* K12 MG1655 and its *relA* deficient mutant, which has its stringent response disabled *(60)*. Genes that were differentially regulated between the two strains over the 30-minute time-course were determined as follows. Linear regression was used to fit the parameters $\boldsymbol{m}$ and $\boldsymbol{b}$ in $\boldsymbol{a} = \boldsymbol{m}t + \boldsymbol{b}$, where $\boldsymbol{a}$ is the gene expression vector and $t$ is time. The slope parameters $\boldsymbol{m}_{WT}$ and $\boldsymbol{m}_{\Delta relA}$ characterize the rate of transcriptional change in each gene. Then, the difference $\langle \boldsymbol{m}_{WT} \rangle - \langle \boldsymbol{m}_{\Delta relA} \rangle$ characterizes the difference in gene regulation between strains having and lacking the stringent response. Let the standard error of $\langle \boldsymbol{m}_{WT} \rangle$ and $\langle \boldsymbol{m}_{\Delta relA} \rangle$ be $\langle\langle \boldsymbol{m}_{WT} \rangle\rangle$ and $\langle\langle \boldsymbol{m}_{\Delta relA} \rangle\rangle$, respectively. Then, the quantity $(\langle \boldsymbol{m}_{WT} \rangle - \langle \boldsymbol{m}_{\Delta relA} \rangle)/\sqrt{\langle\langle \boldsymbol{m}_{WT} \rangle\rangle^2 + \langle\langle \boldsymbol{m}_{\Delta relA} \rangle\rangle^2}$ is normally distributed about zero with unit variance, allowing application of a Z-test. This test results in 183 genes that are differentially regulated (117 up-regulated and 66 down-regulated) between the two conditions (FDR=5%).

The stringent response-regulated genes are quantified as a vector $\boldsymbol{v}$ indexed by the genes, $g$, with $v_g = 1$ if gene $g$ is upregulated, –1 if it is downregulated and 0 otherwise. In our data, each transcriptional response to the adaptive evolution mutations is $\Delta_{gs} = \log_2(\boldsymbol{a}_s/\boldsymbol{a}_{Ref})$, where $s$ is an index over Mut strains and $\boldsymbol{a}$ is gene expression, as before. We define the stringent response score, $S$, for a given Mut strain to be $S = \boldsymbol{\Delta}_s \cdot \boldsymbol{v}$. Statistical significance of the scores was assessed by bootstrapping, in which transcriptional responses were shuffled according to their average expression level across both conditions. The elements of $\boldsymbol{\Delta}_s$ were shuffled (with $s$ held fixed) and $S'$, the randomized stringent response score, was calculated 10,000 times. Using $\langle S' \rangle$ and $\langle\langle S' \rangle\rangle$ to denote the mean and standard deviation of $S'$, respectively, the quantity $(S - \langle S' \rangle)/\langle\langle S' \rangle\rangle$ is normally distributed about zero with unit variance. Applying the Z-test yields the statistical significance of the score against a null model that takes into account the total amount of transcriptional change across all genes in a given Mut strain.

### Origon analysis

Origons are groups of genes reachable from a single master regulator in a transcriptional regulatory network, where master regulators are defined as transcription factors that have no regulatory inputs from other transcription factors *(61)*. Using RegulonDB, we found 82 master regulators and calculated the degree to which they are turned on or off. Let $G$ be a subgraph of the regulatory network reachable from master regulator $O$, and let $x_w$ be the $\log_2$ fold changes for all genes $w$. Define the sign of each edge, $s(u,v)$, to be 1 if $u$ is an activator of $v$, $-1$ if $u$ is a repressor of $v$, and $\text{sign}(x_u)$ otherwise, and further define an indexing function $h(s(u,v)) = 1 + (1 - I(s(u,v) > 0))$, where $I()$ is the indicator function which is 1 if the argument is true and 0 if false. We are now in a position to calculate $y_{wz}$, the contributions to the activation of $O$ for each node $w$, recursively. If $w = O$, the contribution is $y_{O1} = 1$ and $y_{O2} = 0$, since expression of the regulator should contribute to evidence of its activation. For $w \neq O$, let $P(w)$ be the set of direct predecessor nodes of $w$, and define $|P(w)|$ to be the number of predecessors. Then, $y_{w\,h(s(u,v))} = \exp\left(\frac{1}{|P(w)|} \sum_{u \in P(w)} \log\left(r\, y_{u\,h(s(u,v))}\right)\right)$, where $0 < r \leq 1$ is a diffusion parameter to attenuate the

contribution of longer paths. We calculate $y$ starting at $O$, proceeding to its immediate descendants and so on through the network. At each stage, we ensure that all $u \in P(w)$ have been calculated before calculating $y_w$. In the case of cycles, there exists a $u \in P(w)$ such that $u \in D(w)$ where $D(w)$ is the set of descendants (direct and indirect) of $w$. In this case, the we ignore the contribution from all edges $(v, z)$, where $z$ is the ancestor of $u$ and a descendant of $w$, but $v$ is only a descendant of $w$. The averaged contributions to the origon expression is $\overline{y_w} = y_{w1} - y_{w2}$, and the total origon expression is $\overline{y} \cdot x$.

To compare the origon expression against the random expectation, we reshuffle the nodes in the transcriptional regulatory network according to their in-degree and out-degree. We grouped by each value of in-degree from 0 to 12 and binned nodes with in-degrees higher than 12 together (for a total of 14 bins), and we grouped out-degrees into 24 logarithmically spaced groups. Of the possible 336 bins 82 are occupied. To randomize, expression values were randomly permuted between nodes in a given bin, thus preserving the degree distributions under randomization. The results of the origon analysis are reported in Table S7.

### Modulon analysis
Modulons are groups of genes that contribute to a particular cell function. We projected our transcriptional $\log_2$ fold changes onto gene loadings for 92 modulons *(62)*, and compared the resulting projections with those generated by reshuffling the transcriptional data. Modulons were considered significantly differentially expressed if they passed a 1% FDR threshold. Differentially expressed modulons are reported in Table S8.

### RNA alignments of rRNA genes
To examine the RNA-Seq data for signs of transcriptional termination and pausing, we used the wiggle track formatted (WIG) files generated by Rockhopper and the gene locations from NC_000913.3 to calculate a rolling sum (50 bp window) of sequence alignments throughout the genome for both strands. These were averaged over three replicates to generate an alignment profile for each experimental condition. Next, we averaged the alignments for each rRNA operon over three segments: from the start of the *rrs* gene to the start of the tRNA genes, from the start of the tRNA genes to the start of the *rrl* gene, and from the start of the *rrl* gene to the end of the *rrf* gene. Because the tRNA genes are of different lengths in each operon (from 335–447 bp), we rescaled the counts by 447/L, where L is the operon length of the tRNA genes, and linearly interpolated the counts at each base before averaging. Once averaged, these three segments were concatenated into a single vector and annotated with pause sites *(63)* and termination sites *(63, 64, 59)*.

## Results
### Selection of mutations in multiple conditions and their growth consequences
To identify antagonistic gene-environment interactions, we undertook a two-step laboratory evolution approach in which we evolved the WT strain in a defined medium condition and subsequently introduced a sublethal antibiotic stressor (Fig. 1 *A* and Fig. S1 in the Supporting Material). First, we serially passaged the WT strain in M9G, thereby selecting for mutants with enhanced growth rate in this defined condition. Two Ref strains were sequenced revealing a T1037P mutation in *rpoB*, which encodes the β subunit of RNA polymerase (RNAP) or a C8Y mutation in *pykF*, which encodes pyruvate kinase (Fig. 1 *B*). These mutations are known to enhance fitness in anaerobic conditions *(65)* and defined media *(66)*, respectively. Then, each Ref strain was serially passaged in M9G + rif at a concentration that attenuates growth of both strains (Fig. 2). During serial passaging rif-adapted mutants arose, increasing the overall growth rate of the bulk culture and restoring it to pre-rif levels. We isolated a set of 10 independent rif-adapted Mut strains, five from each Ref strain. The growth outcomes following each step of our lab evolution protocol are shown schematically in Fig. 1 *C* and Fig. S1. In the context of our experiments, it is the relative growth rate in M9G between the Mut strains and the Ref strain that determines whether the genotype constitutes an extreme antagonistic pair.

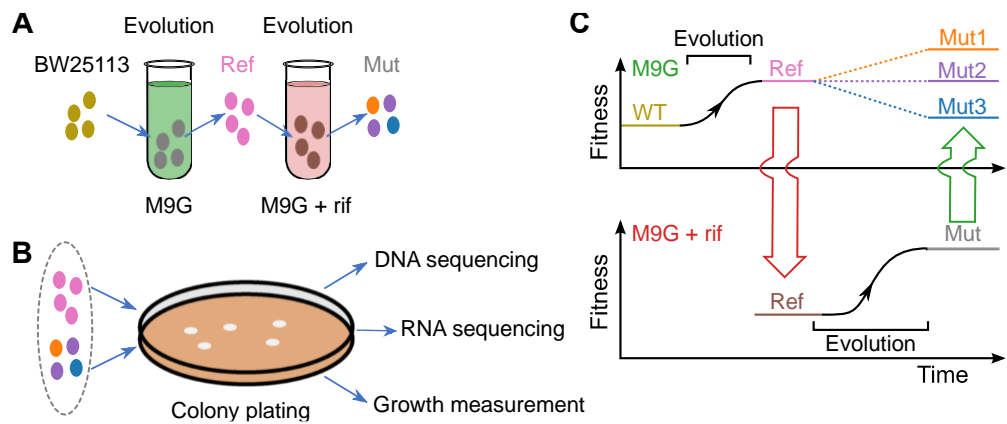

**Figure 1. Experimental approach to identify extreme antagonistic interactions.** (*A*), Diagram of our experimental evolution approach, where WT *E. coli* BW25113 was adaptively evolved in M9G, yielding two Ref strains with an enhanced growth rate. These Ref strains were then evolved in M9G + rif, yielding five rif-adapted Mut strains per Ref parent. (*B*), Analysis of rif-adapted Mut strains and their Ref parents. Selected colonies were analyzed for growth in the presence and absence of rif, were subjected to WGS to identify mutations associated with adaptation, and were RNA-sequenced to detect changes in transcript abundance. (*C*), Fitness of the mutants at each stage of the workflow, as measured by growth rate. The growth rate of the rif-adapted strains were compared with that of their Ref parents: Mut strains with significantly slower growth rate in the absence of rif meet the conditions for extreme antagonistic interactions with the addition of rif (blue), while Mut strains with growth rate equal (purple) or greater (orange) do not.

The genetic changes in each Mut strain are reported in Table 1. Rif binds at the active site of RNAP and inhibits transcription by preventing translocation *(67)*. We note that the mutations conferring enhanced growth rate in M9G found in the Ref strains—*pykF*(C8Y) and *rpoB*(T1037P)—are distinct from those conferring rif resistance in the Mut strains, even though one Ref mutation is in *rpoB*. Genetic changes that the Mut strains acquired during the process of lab evolution in rif all occurred in *rpoB* and cluster at three previously described *E. coli rpoB* rif-resistance sites: (I) the RNAP active site where rif binds, (II) the fork domain which ensures base pair fidelity *(68)*, and (N) the DNA entry channel *(69)*. Mutations at sites I and II are reported to effect transcription slippage *(68)*, those at site II can alter termination *(70)*, while mutations at the N site decrease the open complex lifetime *(69)*. On average, strains harboring mutations at site I have the fastest growth, followed by those at site II and site N. Since all mutations conferring rif adaptation occur in the same gene, we refer to these strains simply by their amino acid substitutions in *rpoB* from this point forward.

| Parent mutation (Ref) | rpoB residue substitution (Mut) | Rif-resistance cluster (67) | Doubling time M9G (min) | Doubling time M9G + rif (min) |
|---|---|---|---|---|
| *pykF*(C8Y) | – | – | 57 | 120 |
| | Q148L | N | 60 | 70 |
| | L511R | I | 63 | 70 |
| | D516G | I | 50 | 57 |
| | H526Y | I | 49 | 53 |
| | I572S | II | 56 | 67 |
| *rpoB*(T1037P) | – | – | 51 | 162 |
| | Q148L | N | 57 | 73 |
| | S508P | I | 47 | 68 |
| | L511R | I | 54 | 65 |
| | I572N | II | 52 | 84 |
| | I572F | II | 55 | 62 |

**Table 1. Mutations and average doubling times for *E. coli* strains grown on M9G in the presence and absence of rif.**

To evaluate whether the mutations associated with rif resistance result in fitness costs in the original cultivation condition (i.e. M9G without rif), we categorize the Mut strains by their growth relative to their Ref parents in M9G. The Mut strains may grow: 1) faster, revealing that mutations are beneficial in both conditions, possibly reflecting incomplete adaptation of the Ref strains to M9G; 2) equally fast, revealing that the mutations are neutral in the original condition; or 3) slower, revealing that the acquired mutations are deleterious in the first condition, thereby indicating antagonism to rif treatment. These three groups are indicated respectively by orange, purple, and blue colors in the subsequent figures and tables, such as in Fig. 2, which presents the growth rates of each strain, and Table S1, which presents doubling times.

We identified rif-adaptive mutations in each of the growth categories in Fig. 2. The fast-growth mutations consist of the uncharacterized, but previously observed *(71)*, S508P mutation (derived from the *rpoB*(T1037P) parent), and the well-described mutations *(72, 67, 73, 68)* at the RNAP active site, D516G and H526Y (derived from the *pykF*(C8Y) parent). Compared to their Ref parent, these mutations boost growth rate by 8%, 15%, and 16%, respectively. While the set of mutations we observed are not exhaustive, growth-enhancing active site *rpoB* mutations were not observed in the *rpoB*(T1037P) Ref strain. The neutral growth mutations—L511R and I572N in *rpoB*(T1037P) and I572S in *pykF*(C8Y)—exhibit few transcriptional similarities, and the existing similarities mostly occur between the Mut strains deriving from the same parent (Table S9 in the Supporting Material).

### Correlations between transcriptional responses to the rif-adaptive mutations

To define the transcriptional changes associated with the mutation combinations in M9G with and without rif, we measured transcript levels in all Ref and Mut strains using RNA-Seq, and determined which genes were differentially expressed between each Mut strain and its Ref parent in both M9G and M9G + rif. We calculated the $\log_2$ fold changes in each case and the Pearson correlations between the fold changes in all genes for each pair of conditions.

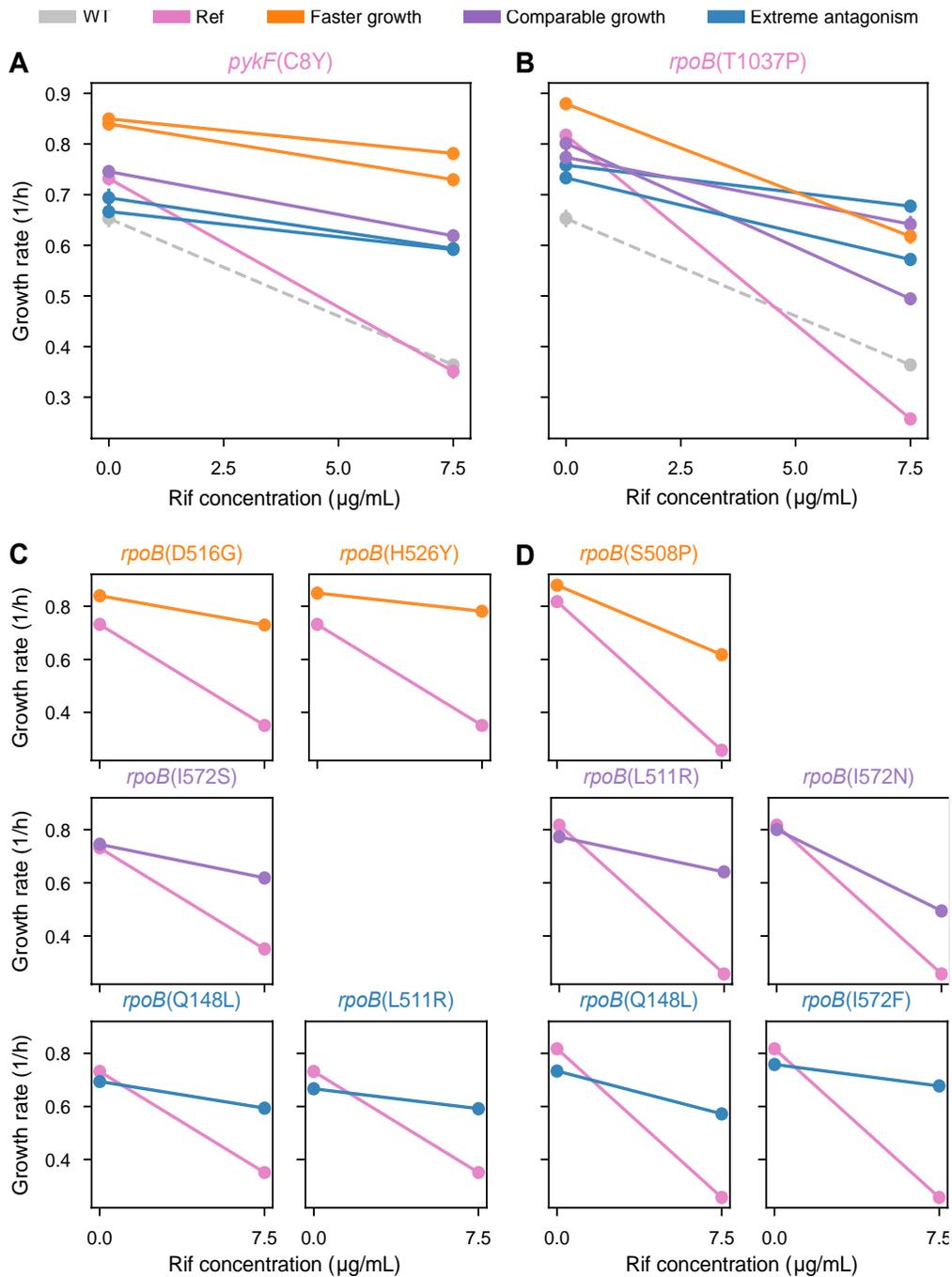

**Figure 2. Growth rates of the WT, Ref, and Mut strains in the presence and absence of rif.** (*A*) and (*B*), Growth rates of the WT strain (gray), Ref strains (pink), and Mut strains (other colors) derived from the *pykF*(C8Y) parent (*A*) and the *rpoB*(T1037P) parent (*B*). Mut strains are color-coded according to whether they grow faster than (orange), as fast as (purple), or slower than (blue) the Ref strain in the absence of rif. (*C*) and (*D*), Growth rate comparisons of the Mut strains with the *pykF*(C8Y) parent (*C*) and the *rpoB*(T1037P) parent (*D*) in the presence and absence of rif. The means are denoted by circles and the standard errors of the mean are denoted by error bars, which are smaller than the size of the circle in most cases.

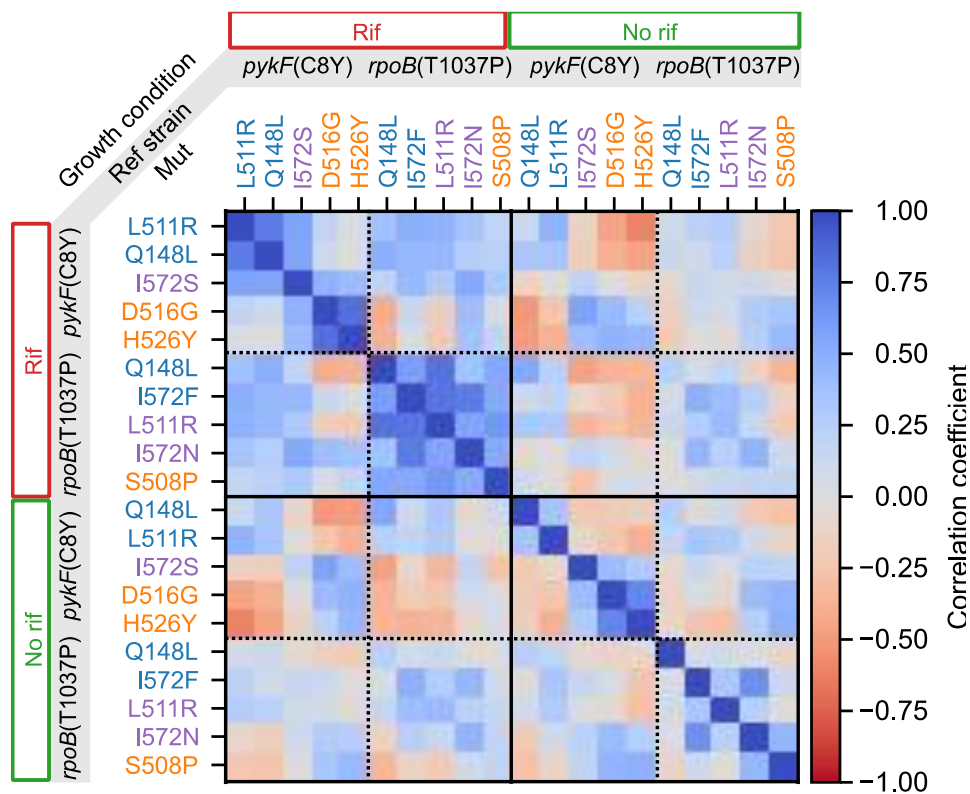

**Figure 3. Transcriptional correlations between the Mut strains in M9G and M9G + rif.** The color code indicates the correlations between $\log_2$ fold changes for each strain in each cultivation condition. The colors of the labels indicate the growth of each Mut strain relative to its Ref parent as defined in Fig. 1

There are five major trends in the transcriptional correlations shown in Fig. 3. Specifically: (i) all Mut strains are more highly correlated in the presence of rif than in its absence; (ii) the mutations at the I572 residue are all correlated with I572F and I572N being strongly so; (iii) the five fastest-growing Mut strains (relative to their Ref parents), S508P, D516G, H526Y, I572N, and I572S, are correlated in the presence and absence of rif; (iv) the Q148L and L511R Mut strains derived from different Ref parents have strongly correlated transcriptomes; and (v) all the slow-growing Q148L and L511R Mut transcriptomes are anticorrelated with those of the fast-growing H526Y, D516G, and S508P strains in both conditions. Finding (i) suggests that all Mut strains are in similar biochemical/physiologic space when dealing with the challenge of rif, while finding (ii) raises the question of how much the Ref strain mutation determines the transcriptional response versus the chemical similarity of the substituted amino acid. Together, findings (iii), (iv), and (v) imply that mutants that grow faster than their parent Ref converge on a metabolic/physiologic profile that is distinct from mutants with slower growth. The correlation values associated with Fig. 3 are included in Table S9.

### Patterns in the annotations of differentially expressed genes

To discover functional patterns at the level of gene regulation, we performed annotation analysis using the PANTHER annotations *(56)*, COG annotations *(57)*, modulons *(62)*, and origons *(61)* from RegulonDB *(59)*. Of the four annotation schemes we consider, the origon analysis is the most fine-grained because it links transcription factors directly with the regulated genes.

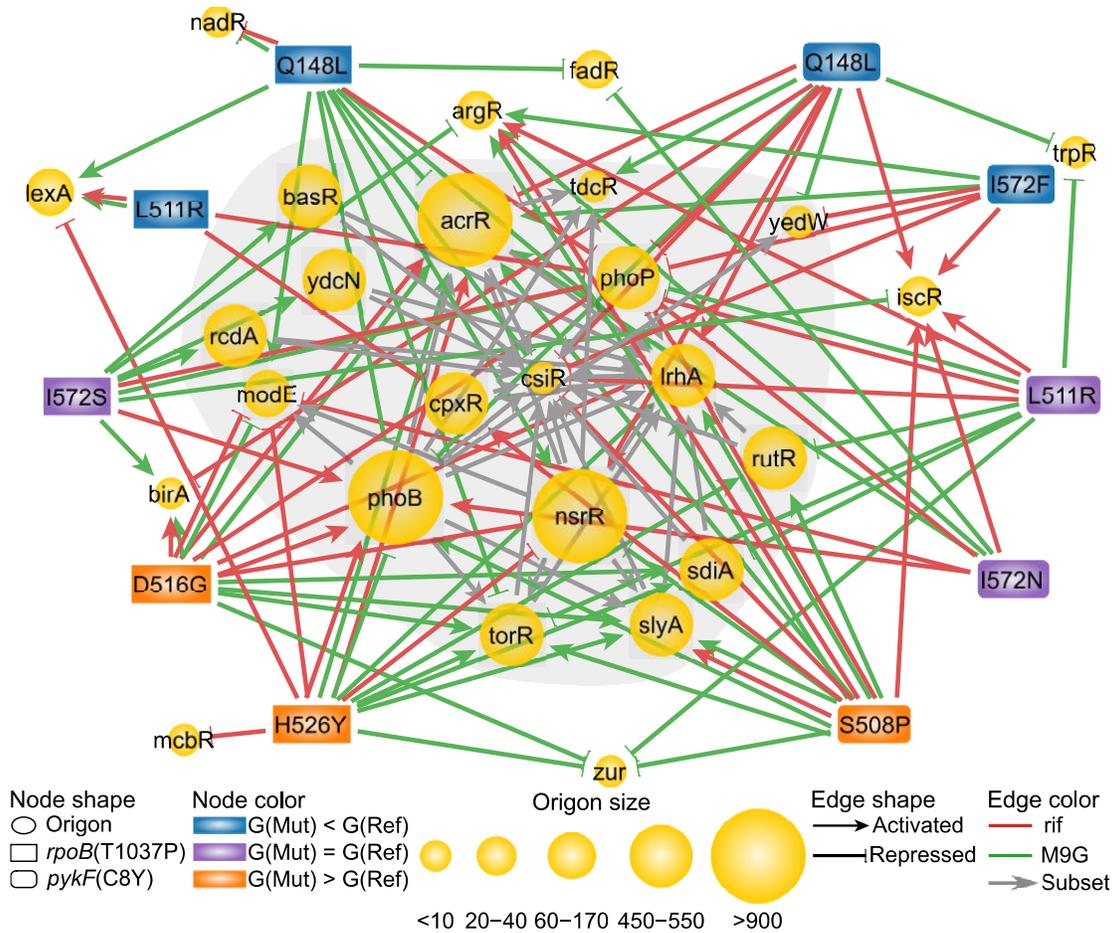

**Figure 4. Network of significant origon expression changes in the Mut strains relative to Ref parents.** The node color, node shape, node size, edge color, and arrow shape are defined in the legend, where G(Mut) denotes the growth rate of the Mut strain in M9G. Origons are labeled by their master regulator, and all origons that have a subset relationship with another are placed on the gray background.

We visualize the origon results in Fig. 4, using a network representation. There are two types of network edges: those that connect Mut strains with their differentially expressed origons in M9G and rif and those that connect two origons. The latter edge type indicates that the genes downstream of the master regulator in the origon at the edge tail are contained within the larger origon at the edge head. When considering genes included in origons, we observe all strains except H526Y share *phoP* downregulation in rif, while *rpoB*(T1037P)-derived strains also share *iscR* upregulation. In the absence of rif, the cluster I fast-growth strains share *zur* downregulation and *acrR*, *torR*, and *slyA* upregulation, while the I572F and I572N strains share *acrR* and *argR* upregulation, and the Q148L strains downregulate *phoB*.

The origons that are differentially expressed are generally all implicated in *E. coli*'s stress response, a trend corroborated by our analysis of modulons. The membrane, GadEWX, and His-tRNA modulons, the latter two of which are implicated in acid stress response and histidine metabolism, respectively, tend to be upregulated in the presence of rif in all strains. Simultaneously, iron utilization and CysB, which are implicated in inorganic sulfate utilization, are generally downregulated. Overall the results in rif exhibit the convergence in transcription observed in Fig. 3, that is, the transcription factors that Mut strains express in M9G + rif are more similar than those expressed in M9G (Tables S2-S8). In the absence of rif, the modulon activation for slow-growth strains show upregulation of RpoS and GadEWX, which is generally anticorrelated with that of fast-growth strains. The anticorrelation is particularly strong in the *pykF*(C8Y) background, whereas the Q148L and I572F strains show few similarities with each other despite their similar growth phenotype.

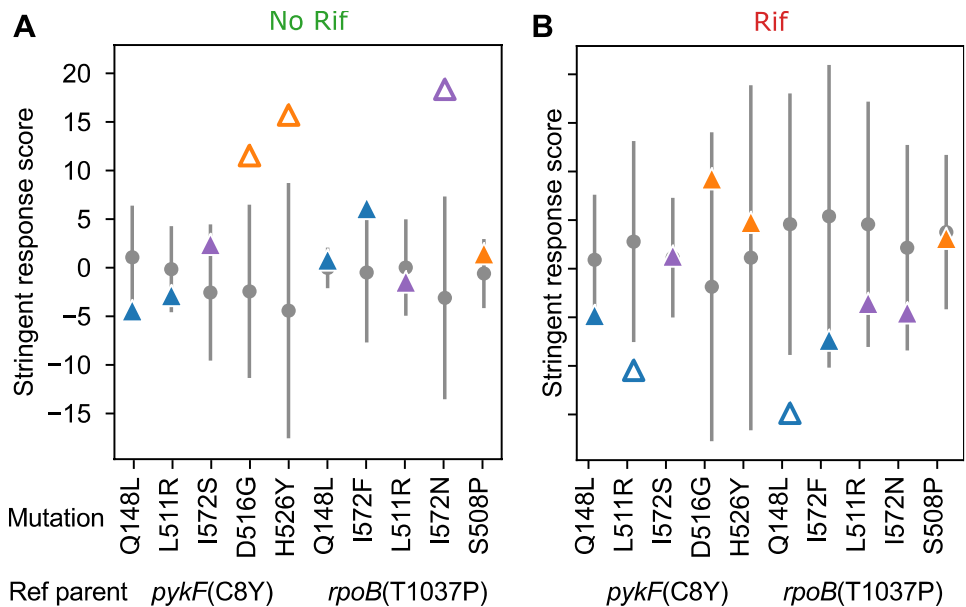

**Figure 5. Stringent response score for each Mut strain.** (*A*) and (*B*), Comparison between the observed stringent response scores (triangles) and bootstrapped means (circles) in M9G (*A*) and M9G + rif (*B*). Statistically significant cases are denoted by the unfilled symbols, where the error bars correspond to the 95% confidence interval determined by bootstrapping. The color code indicates the Mut strain's growth rate relative to the Ref parent as defined in Fig. 1.

As several of the most strongly differentially expressed modulons are in stress response genes, we analyze genes in the stringent response regulon in Fig. 5, given that genes in this regulon were implicated in the evolution of enhanced growth rate in previous experiments *(69, 19)*. The *E. coli* stringent response downregulates genes involved in ribosome synthesis and upregulates genes for amino acid biosynthesis to account for poor nutrient conditions *(74)*. For each transcriptional profile, we calculated a stringent response score by counting the number of genes that change in accordance with their behavior during the stringent response (see Materials and Methods). The fastest growing Mut strains (D516G, H526Y, and S508P) have transcriptional profiles that resemble the stringent response in the absence of rif, while the slower growing strains appear to oppose, or not exhibit, this response. This pattern is particularly pronounced in the *pykF*(C8Y) background but is not apparent in the *rpoB*(T1037P) background.

As expected, strains that have a stringent-like transcriptional profile also have decreased transcripts corresponding to ribosomal proteins and 5S rRNA (Fig. S2 in the Supporting Material). These strains show down-regulation of genes classified in translation, nucleotide synthesis, and amino acid synthesis COG groups, and exhibit *fnr* activation and significantly enhanced expression of glycolytic genes *eno*, *pgk*, *pykF*, and *tpiA*. Together, these trends indicate a shift toward overflow metabolism in the faster-growing strains, wherein the majority of ATP production is shifted from oxidative phosphorylation to glycolysis *(75)*. Meanwhile, transcriptional changes in fast-growing strains are anticorrelated with those observed in the slow-growth mutants L511R and Q148L— including three of the four extreme antagonistic cases—possibly reflecting a maintenance of oxidative metabolism, which is more energy efficient (Fig. 5 and Tables S2-S8). Thus, slow-growth and fast-growth mutations appear to restore growth through different means, and transcriptional changes that emulate those observed in the stringent response are associated with fast growth in the *pykF*(C8Y) background.

We analyzed the rRNA and ribosomal protein transcript data for all strains (Fig. S2). Rif-adapted mutants harboring I572 mutations and those growing faster than their parent in the absence of rif exhibit decreased amounts of ribosomal protein transcripts in that condition. Increased tRNA and 5S rRNA expression may indicate increased readthrough from more upstream genes in the rRNA locus; decreases of particular genes at the rRNA locus may likewise provide evidence for transcription termination. Elevated expression of amino acid biosynthetic genes may reflect a short-lived open complex. We note that changes in rRNA may reflect an imbalance in ribosomal components, analogous to suboptimal protein to DNA ratios observed in previous extreme antagonistic drug interactions generally *(41)* and in rif specifically *(76)*.

The varying transcriptional patterns of genes encoding ribosomal components motivated us to more closely examine the reads that mapped to the rRNA loci. This effort was aimed at determining whether the mutant RNAPs exhibited differences in transcriptional efficiency along the operon. In Fig. 6 *A*, we plot the fold change in read alignment averaged across all rRNA operons for cells cultivated with and without rif. In the presence of rif, we see that both Ref strains have a pronounced peak in the 23S rRNA genes, 720 bp downstream of the *rrl* transcriptional start site. We note that the height of the peak correlates with the cost of the Ref mutations in rif as the *rpoB*(T1037P) mutant has slower growth in rif than the *pykF*(C8Y) mutant and a corresponding higher peak. On the antisense strand, we observe a peak immediately downstream of this site. Rif resistance mutations appear to decrease the peak height (Fig. 6 *B*), but the extent to which they do so is correlated with growth only in the *rpoB*(T1037P) mutant derived strains. When cultivated without rif, the change in number of aligning sequences at this site is positively correlated with growth (Fig. 6 *C*). While extreme antagonistic mutations appear to constitutively decrease the number of alignments at this site, the fast growth mutations show substantially decreased alignments only in the presence of rif.

We compared the peak locations to those of pause sites *(63)* and rho-dependent termination sites *(64)*. There is a rho-dependent termination site on the antisense strain 784 bp downstream of the rrl start on the sense strain and a pause site 832 bp downstream of the rrl start proximal to the 720bp peak. In addition, a secondary site 1349 bp downstream from the *rrl* start recapitulates the behavior of the primary peak in response rif treatment. When comparing transcription of both Ref strains to their derived Mut strains in rif, decreases in alignments are apparent, but the ordering of strains is distinct from that observed at the 720 bp site. In the absence of rif, alignments increase by a larger amount in the *pykF*(C8Y) background at this site, which also has a predicted rho-dependent termination site. Beyond these two peaks, the alignment trends show decreases in transcriptional readthrough at the end of genes in fast-growth strains.

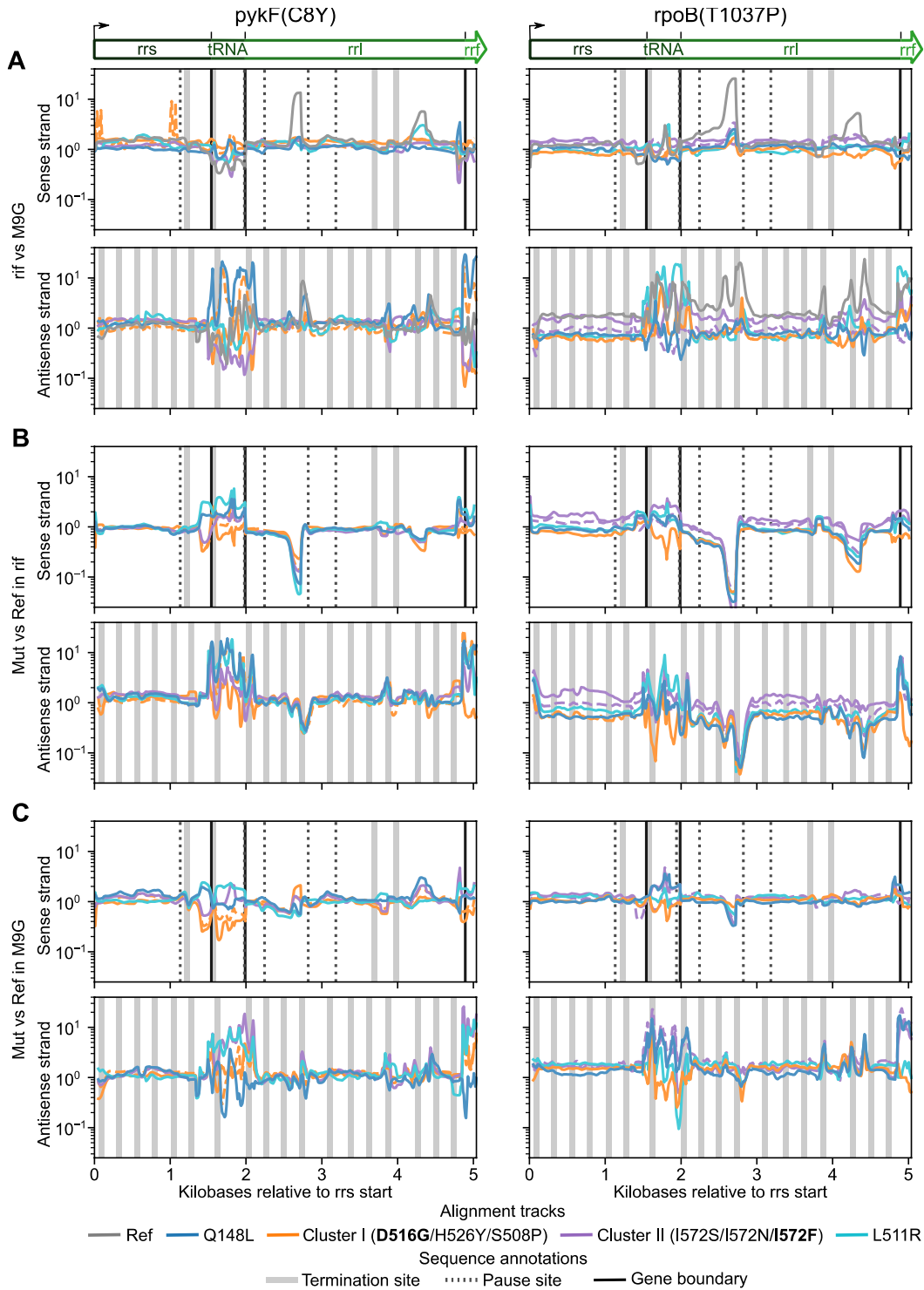

**Figure 6. Fold change in RNA-Seq alignments on both strands averaged across all rRNA operons.** *(A)* Comparison of strains cultivated with and without rif. Data is averaged over the seven rRNA operons, whose structure is illustrated by the cartoon above the panel. The top panel shows the sense strand alignment and the bottom shows the antisense alignment. *(B)* and *(C)* Same as *(A)*, except comparing Mut strains to their Ref parent when cultivated with rif *(B)* or without rif *(C)*. Bold face strains have dashed lines in the figures and legends define the meaning of the remaining line colors, color backgrounds, and line styles.

## Discussion

Our identification of extreme antagonistic interactions reveals the role of molecular mechanisms fomenting gene-environment antagonism, including connections with transcriptional pausing and the stringent response. The application to an antibiotic environment reveals that the rif-adaptive evolution mutations alter the kinetic and thermodynamic properties of RNAP to counter the drug action but hamper growth in the drug's absence—consistent with the pattern of resistance mutations being pleiotropic *(49, 48)* and suboptimal *(41, 43)*. Our study also demonstrates a proof-of-concept approach for the discovery of gene-environment interactions that is scalable to multiple genetic and environmental conditions and sufficiently sensitive to identify extreme antagonistic interactions. A central aspect of our approach is the two steps of adaptive evolution, as schematized in Fig. S1. The first step, pre-adaptation to the initial environmental condition prior to applying the environmental perturbation, facilitates the identification of extreme antagonistic interactions *(77, 78)*. The second step, adaptation to an antibiotic environment, reveals mutations involved in extreme antagonistic interactions that (as further discussed below) constitute candidate targets for the development of new antibiotic combinations.

Pre-adaptation to a defined medium reduces the likelihood that adaptive mutations to antibiotic stress would enhance fitness in the absence of an antibiotic, as predicted by Fisher's geometric model *(79, 80)*. This is a step that was only recently explored in *Pseudomonas aeruginosa (12)*, and our transcriptional analysis extends beyond this effort. We distinguish our pre-adapted strains, which are kept in exponential phase, from strains deriving from the long-term evolution experiment in *E. coli* B, in which cells transition daily between exponential and stationary phase *(66, 77)* and from unadapted strains in previous studies focused on non-extreme antagonistic interactions *(81, 33, 82)*. The impact of pre-adaptation is illustrated in Fig. 2 *A* and *B* by the intersecting lines, which show that M9G-adaptive mutations can sensitize strains to the addition of rif. This finding stands in contrast with studies investigating random mutations *(83, 84)*, collateral drug resistance *(81)*, and nutrient changes *(85, 86)*, which suggest that adaptive mutations desensitize strains to other stresses. Indeed, direct selection of mutations conferring antibiotic resistance from unadapted WT strains tends to yield mutations that are beneficial in both environments *(87-89)*, including the D516G and H526Y mutations in *E. coli* cultivated in M9G *(19)*. We suggest that the difference originates from a higher cost of selection *(79)* across environments of a different nature, which in the case of our experiment implies that the same mutation generally cannot provide fast growth in both environments (with and without rif). The compatibility between these contrasting results can indeed be appreciated by recognizing that previous studies have mainly compared the response to environments of the same nature, such as different antibiotics *(81, 43, 12)* or carbon sources *(85, 86)*. Indeed, antibiotic stress causes a targeted impairment of the cell's biochemical network that does not necessarily have an equivalent in adaptation to a nutrient condition. Our demonstration that adaptation to a non-antibiotic condition sensitizes the organism to antibiotic stress is corroborated by recent experiments in *Enterococcus faecalis (90)*, and in accordance with observations of non-reciprocating collateral sensitivity between carbenicillin and gentamycin *(12)*, we anticipate that the order of adaptive steps matters.

A possible mechanism explaining this tradeoff between generalization and specialization is a difference in transcriptional dynamics *(70, 91, 92)*. Generally, the transcriptional changes observed between Ref and Mut strains in the presence of rif highlight the transcriptome-wide impact of adaptation while the changes in the absence of rif reflect the cost of resistance. The alignment changes in Fig. 6 suggest that termination could play a role in rif resistance, consistent with the observed changes in termination caused by mutations at the D516, H526, and I572 residues of *rpoB (70)*. These findings may relate to rif resistance in *M. tuberculosis (50),* as recent work in that organism has implicated termination efficiency as the mechanism underlying the cost of resistance *(51)*.

The experimental requirements to identify extreme antagonistic interactions are arguably more involved than those required to identify synergistic and merely antagonistic ones. While we expect extreme antagonistic interactions to be large in number, they are also expected to be comparatively rare in the context of all interactions. This renders *reverse* genetic screening approaches, such as those based on libraries of gene knockouts *(2-6)*, impractical due to combinatorial explosion. Our *forward* genetic approach identifies point mutations that have pleiotropic effects on protein function in essential genes, which is a task that falls outside the scope of typical high-throughput approaches, and it is potentially massively parallelizable given advances in continuous culture and reductions in the cost of sequencing. Although high-throughput knockout experiments are not easily scalable in conjunction with pre-adaptation, future implementations of our approach could involve hybrid forward and reverse genetic approaches combined with mutagenesis techniques like TnSeq *(5, 93)* to expand the range of possible extreme antagonistic responses.

The effects of mutations in extreme antagonistic interactions isolated in the second adaptive step constitute candidates for the development of antagonistic antibiotic partners for rif. This proposition is motivated by previous demonstrations that extreme antagonistically interacting drugs—in which the combination of two drugs is weaker than one of the drugs alone—can select against cells resistant to the suppressor drug *(44, 45, 32)*. The mutations we identified can thus be regarded as targets for the design of one such partner for rif, since a drug that emulates these mutations would suppress growth in M9G (thus acting as an antibiotic) and would alleviate growth defects in rif media (thus exhibiting an extreme antagonistic interaction with rif). In Fig. 2 *C* and *D* (bottom row), the region of antagonistic interactions corresponds to rif concentrations on the right side of the intersection between the lines. For example, for a rif partner corresponding to the Q148L mutation in *rpoB*(T1037P), resistance would reduce growth rate from 0.7 $h^{-1}$ to 0.38 $h^{-1}$ when the concentration of rif is 7.5 µM. The process of target identification for the design of drug partners can be further optimized by using error-prone polymerases *(94)* and other approaches that generate variability by manipulating selective pressure *(95, 96)*. Alternatively, after target genes have been characterized partner drugs may be directly identified by screening small molecules that act on the target gene. We note that the rif-adaptive mutations in our experiments occur in *rpoB*, suggesting that RNAP will be the likely target of antagonistic partners of rif that can be developed to select against resistance.

Finally, while here we have focused on pairwise gene-environment interactions, this study can also inform the development of methods to optimize the scheduling of drug treatments *(43, 33, 36)*, and to identify higher order antagonistic interactions *(47, 97)*, in which the antagonism emerges between three or more genes and/or drugs. We anticipate that applications and generalizations of our approach will contribute to countering antibiotic resistance and advance our understanding of the role of antagonism in cellular networks *(98)* of multiple organisms.

## Author Contributions

TPW, AJ, SC, and AEM designed the research. MZ and AF performed the evolution experiments and mutational mapping analysis. TPW analyzed and interpreted the transcriptomic and growth data. TPW led the writing of the manuscript with contributions from all authors. All authors approved the final manuscript.

## Acknowledgements

**Funding:** This research was supported by NIH/NIGMS R01 GM113238.

**Competing Interests:** The authors declare that no competing interests exist.

**Data and Materials Availability:** All data necessary for the reproduction of the reported results is included in the Supporting Material. Materials used to obtain the results and computer code for performing the analysis is available from the authors upon reasonable request. Raw sequencing data have been deposited in the NCBI Sequence Read Archive under the submission number SRP220530 (BioProject PRJNA564142).

## Contents of Supporting Material
**Supporting Material**
    Tables S1–S9
    Figures S1–S2

# Supporting Material

**SUPPLEMENTAL FIGURES**

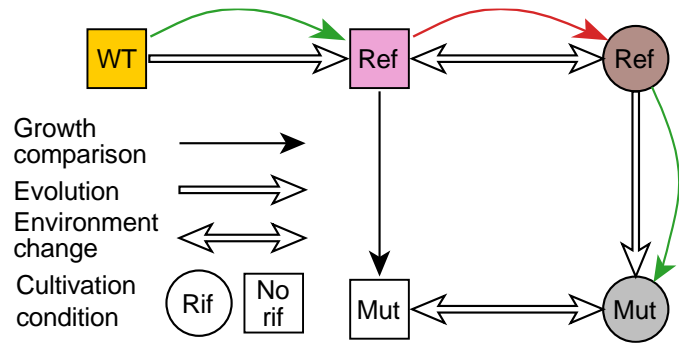

**Figure S1. Schematic of the assay to detect extreme antagonistic interactions.** The WT strains were adaptively evolved in M9G medium, resulting in Ref strains. The Ref strains were shifted to an antibiotic medium condition, which was treatment with 7.5 µg/mL rif. Fast-growing, rif adapted, mutants were subsequently evolved, yielding Mut strains. The growth rates of the Ref and Mut strains were measured in both the original (M9G) and stressful (M9G + rif) media. In the diagram, evolutionary changes are represented by single-headed arrows and environmental shifts are represented by double-headed arrows. Green and red edges indicate increases and decreases in growth rate, respectively. The node colors match their corresponding strains in Fig. 1 *C*.

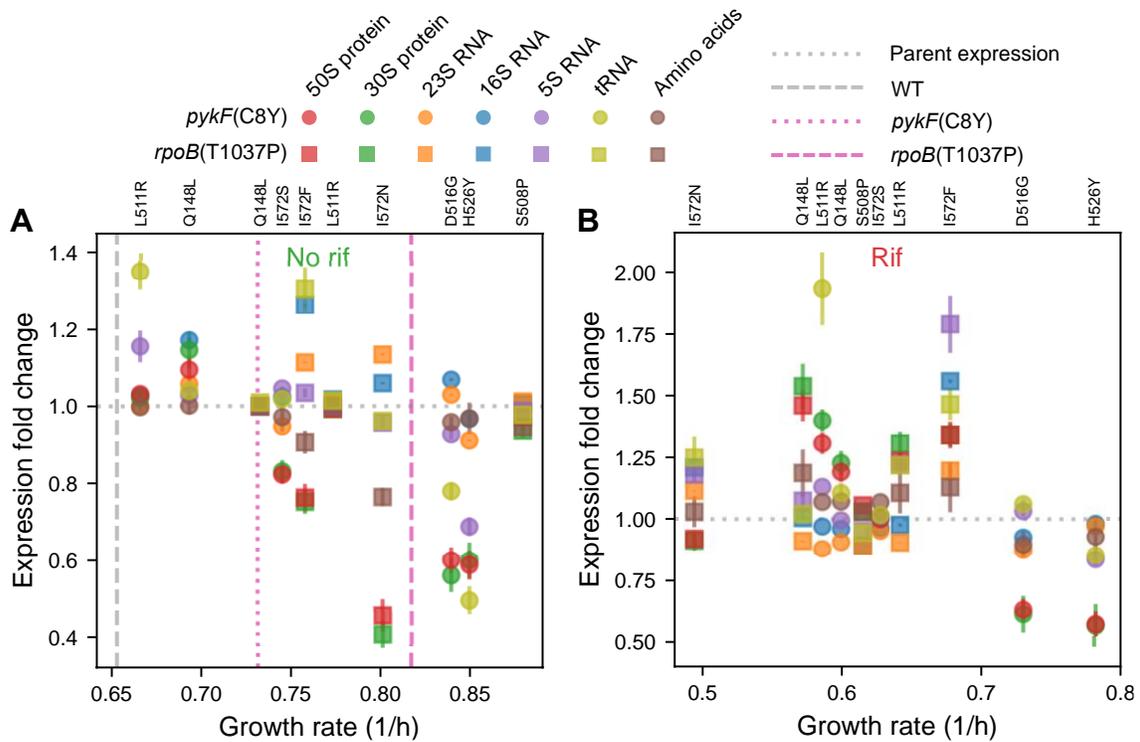

**Figure S2. Expression of rRNA loci and transcripts of ribosomal proteins as a function of growth rate.** (*A*) and (*B*), Expression fold change for strains cultivated in M9G (*A*) and M9G + rif (*B*), where the error bars indicate the standard errors of the mean. The symbols encode the Ref parent strain and the color indicates the transcript identity for a given Mut strain, as indicated by the corresponding mutation (top text). Wild type and parent growth rates in M9G are indicated by vertical lines and the parental expression is indicated by the horizontal dotted line. The growth rates of sensitive strains are omitted in (*B*) to show the Mut strain growth rates in more detail.

# SUPPLEMENTAL TABLES

**Table S1. Raw doubling times used to generate Fig. 2 and statistical analysis characterizing growth of each Mut strain relative to that of the Ref strain.**

| Doubling times in min for BW25113, pykF(C8Y) and rpoB(T1037P) reference strains, and the rif-adapted strains grown in M9G. | | | | | | | |
|---|---|---|---|---|---|---|---|
| **M9G Adaptive Mutation** | **Rif Adaptive Mutation** | **Rep1** | **Rep2** | **Rep3** | **Rep4** | ***t*-test *p*-value** | ***t*** |
| BW25113 | None | 67 | 66 | 60 | 62 | N/A | N/A |
| pykF(C8Y) | None | 60 | 57 | 54 | 57 | N/A | N/A |
| pykF(C8Y) | rpoB(H526Y) | 48 | 49 | 49 | 50 | 9.53E-03 | -5.94 |
| pykF(C8Y) | rpoB(D516G) | 52 | 50 | 47 | 49 | 5.31E-04 | -16.00 |
| pykF(C8Y) | rpoB(L511R) | 67 | 67 | 57 | 59 | 4.64E-02 | 3.28 |
| pykF(C8Y) | rpoB(I572S) | 55 | 57 | 55 | 55 | 5.50E-01 | -0.67 |
| pykF(C8Y) | rpoB(Q148L) | 64 | 62 | 55 | 60 | 3.96E-02 | 3.50 |
| rpoB(T1037P) | None | 50 | 51 | 51 | 52 | N/A | N/A |
| rpoB(T1037P) | rpoB(S508P) | 46 | 46 | 47 | 49 | 3.58E-03 | -8.37 |
| rpoB(T1037P) | rpoB(L511R) | 51 | 51 | 57 | 57 | 1.48E-01 | 1.94 |
| rpoB(T1037P) | rpoB(I572N) | 56 | 50 | 50 | 51 | 5.36E-01 | 0.70 |
| rpoB(T1037P) | rpoB(I572F) | 55 | 55 | 55 | 54 | 5.54E-03 | 7.19 |
| rpoB(T1037P) | rpoB(Q148L) | 59 | 57 | 57 | 54 | 2.01E-02 | 4.53 |

| Doubling times in min for BW25113, pykF(C8Y) and rpoB(T1037P) reference strains, and the rif-adapted strains grown in M9G with rif. | | | | | | | |
|---|---|---|---|---|---|---|---|
| **M9G Adaptive Mutation** | **Rif Adaptive Mutation** | **Rep1** | **Rep2** | **Rep3** | **Rep4** | ***t*-test *p*-value** | ***t*** |
| BW25113 | None | 117 | 109 | 120 | 113 | N/A | N/A |
| pykF(C8Y) | None | 114 | 138 | 106 | 121 | N/A | N/A |
| pykF(C8Y) | rpoB(H526Y) | 53 | 55 | 53 | 51 | 1.62E-04 | -23.81 |
| pykF(C8Y) | rpoB(D516G) | 58 | 55 | 58 | 58 | 4.95E-04 | -16.39 |
| pykF(C8Y) | rpoB(L511R) | 72 | 72 | 70 | 68 | 1.04E-03 | -12.74 |
| pykF(C8Y) | rpoB(I572S) | 68 | 69 | 63 | 68 | 3.50E-04 | -18.41 |
| pykF(C8Y) | rpoB(Q148L) | 71 | 73 | 67 | 69 | 4.33E-04 | -17.14 |
| rpoB(T1037P) | None | 179 | 156 | 152 | 162 | N/A | N/A |
| rpoB(T1037P) | rpoB(S508P) | 74 | 67 | 63 | 66 | 1.00E-06 | -123.46 |
| rpoB(T1037P) | rpoB(L511R) | 65 | 64 | 60 | 71 | 4.40E-05 | -36.89 |
| rpoB(T1037P) | rpoB(I572N) | 88 | 87 | 81 | 81 | 7.20E-05 | -31.20 |
| rpoB(T1037P) | rpoB(I572F) | 63 | 58 | 64 | 61 | 2.60E-05 | -44.04 |
| rpoB(T1037P) | rpoB(Q148L) | 74 | 78 | 70 | 69 | 8.90E-05 | -29.17 |

**Table S2. Annotation analysis RegulonDB annotations.**

| Media | M9G | | | | | | | | | | Rif | | | | | | | | | |
|---|---|---|---|---|---|---|---|---|---|---|---|---|---|---|---|---|---|---|---|---|
| Ref | pykF(C8Y) | | | | | rpoB(T1037P) | | | | | pykF(C8Y) | | | | | rpoB(T1037P) | | | | |
| Adaptation | D516G | H526Y | L511R | Q148L | I572S | I572F | I572N | L511R | Q148L | S508P | D516G | H526Y | L511R | Q148L | I572S | I572F | I572N | L511R | Q148L | S508P |
| TF Name | | | | | | | | | | | | | | | | | | | | |
| gadE | 0 | -1 | 0 | 0 | 0 | 0 | 0 | 1 | 0 | 0 | 1 | 0 | 1 | 1 | 1 | 1 | 1 | 1 | 1 | 1 |
| dcuR | 1 | 0 | 0 | 0 | 1 | 1 | 0 | 0 | -1 | 1 | 1 | 1 | 0 | 0 | 0 | 0 | 0 | -1 | -1 | 0 |
| gatR_2 | 0 | 0 | 1 | 1 | 0 | -1 | 0 | 0 | 1 | 0 | 0 | -1 | 0 | 0 | -1 | -1 | -1 | 0 | 1 | 0 |
| malT | -1 | 0 | 0 | 0 | -1 | 0 | -1 | 0 | 0 | -1 | 0 | 1 | 1 | 1 | 1 | 0 | 0 | 0 | 0 | 0 |
| birA | 0 | 1 | 0 | -1 | 1 | 0 | 0 | -1 | 0 | 0 | 1 | 0 | 0 | 0 | 0 | -1 | 0 | -1 | -1 | 0 |
| gadX | 0 | -1 | 0 | 0 | 0 | 0 | 1 | 0 | 0 | 0 | 1 | 0 | 1 | 0 | 1 | 1 | 1 | 1 | 0 | 1 |
| argR | 0 | 0 | 0 | 0 | 0 | 1 | 1 | 1 | 0 | 0 | 0 | 0 | -1 | -1 | -1 | 0 | 1 | 0 | 0 | 1 |
| marA | 0 | 0 | 0 | 1 | -1 | 0 | -1 | -1 | 0 | 0 | -1 | 0 | 0 | 0 | -1 | 0 | -1 | 0 | 0 | 0 |
| hprR | 0 | 0 | 0 | -1 | 0 | 0 | 0 | 0 | -1 | 0 | 0 | 0 | 0 | 0 | 0 | -1 | -1 | -1 | -1 | -1 |
| cusR | -1 | -1 | 0 | 0 | 0 | -1 | -1 | 0 | 0 | -1 | 0 | 0 | 0 | 1 | 0 | 0 | 0 | 0 | 0 | 0 |
| fis | -1 | -1 | 1 | 0 | 0 | 0 | -1 | 0 | 0 | 0 | 0 | 0 | 0 | 0 | 0 | 1 | 0 | 0 | 0 | 0 |
| fadR | -1 | -1 | 0 | 0 | -1 | -1 | -1 | 0 | 0 | 0 | 0 | 0 | 0 | 0 | 0 | 0 | 0 | 0 | 0 | 0 |
| arcA | 1 | 0 | 0 | 0 | 0 | 0 | 1 | 0 | 1 | 1 | 1 | 0 | 0 | 0 | 0 | 0 | 0 | 0 | 0 | 0 |
| cpxR | 0 | 0 | 0 | 1 | 0 | 0 | -1 | 0 | 0 | 0 | -1 | 0 | -1 | 0 | -1 | 0 | 0 | 0 | 0 | 0 |
| nhaR | 0 | 0 | 0 | -1 | 0 | 0 | 0 | 0 | 0 | 1 | 1 | 1 | 0 | 0 | 0 | 0 | 0 | 0 | 0 | 1 |
| fur | 0 | 0 | 0 | 0 | 0 | 0 | 0 | 0 | 0 | 0 | 0 | 0 | -1 | -1 | 0 | 0 | 0 | -1 | -1 | -1 |
| cysB | 0 | 1 | 0 | 0 | 0 | 0 | 0 | 0 | 0 | 0 | 0 | 0 | 0 | 0 | 0 | 0 | 0 | -1 | -1 | -1 |
| iscR | 0 | 0 | 0 | 0 | -1 | 0 | 0 | 0 | 0 | 0 | 0 | 0 | 0 | 0 | 0 | 0 | 0 | 1 | 1 | 1 |
| metJ | 0 | 0 | 0 | 0 | 0 | 1 | 0 | 0 | 0 | 1 | 1 | 1 | 0 | 0 | 0 | 0 | 0 | 0 | 0 | 0 |
| soxS | 0 | 0 | 0 | 1 | 0 | -1 | -1 | 0 | 0 | 0 | 0 | 0 | 0 | 0 | 0 | 0 | 0 | 0 | 0 | 0 |
| hns | 0 | 0 | 0 | 0 | 1 | -1 | -1 | 0 | 0 | 0 | 0 | 0 | 0 | 0 | 0 | 0 | 0 | 0 | 0 | 0 |
| cecR | 0 | 0 | 0 | 0 | 0 | -1 | -1 | -1 | 0 | 0 | 0 | 0 | 0 | 0 | 0 | 0 | 0 | 0 | 0 | 0 |
| hypT | 0 | 0 | 0 | 0 | 0 | -1 | 0 | 0 | 0 | 0 | 0 | 0 | 0 | 0 | -1 | 0 | -1 | 0 | 0 | 0 |
| lrp | 0 | 0 | 0 | 0 | 0 | -1 | 0 | 0 | 0 | 0 | 0 | 0 | 0 | 0 | 1 | 0 | -1 | 0 | 0 | 0 |
| fnr | 0 | 0 | 0 | 0 | 0 | 0 | 0 | 0 | 0 | 1 | 1 | 1 | 0 | 0 | 0 | 0 | 0 | 0 | 0 | 0 |
| rstA | 0 | 0 | 0 | 0 | 0 | 0 | 0 | 0 | 0 | 0 | 0 | 0 | 0 | -1 | 0 | 0 | 0 | -1 | -1 | 0 |
| mlrA | -1 | 0 | 0 | 0 | 0 | 0 | 0 | 0 | 0 | 0 | 0 | 0 | 1 | 0 | 0 | 0 | 0 | 0 | 0 | 0 |
| lexA | 0 | 0 | 0 | 0 | 0 | 0 | -1 | 0 | 0 | 0 | 0 | 0 | 0 | 0 | 0 | 0 | -1 | 0 | 0 | 0 |
| pdhR | 0 | 0 | 0 | 0 | 0 | 0 | 0 | -1 | 0 | 0 | 0 | 0 | 0 | 0 | 1 | 0 | 0 | 0 | 0 | 0 |
| phoP | 0 | 0 | 0 | 0 | 0 | 0 | 0 | 1 | 0 | 0 | 0 | 0 | 0 | 0 | 0 | 0 | 1 | 0 | 0 | 0 |
| csiR | 0 | 0 | 0 | 0 | 0 | 0 | 0 | 0 | 0 | 1 | 0 | 0 | 0 | -1 | 0 | 0 | 0 | 0 | 0 | 0 |
| nagC | 0 | 0 | 0 | 0 | 0 | 0 | 0 | 0 | 0 | 0 | 0 | 0 | 0 | -1 | 0 | 0 | 1 | 0 | 0 | 0 |
| mraZ | 0 | 0 | 1 | 0 | 0 | 0 | 0 | 0 | 0 | 0 | 0 | 0 | 0 | 0 | 0 | 0 | 0 | 0 | 0 | 0 |
| nac | 0 | 0 | 1 | 0 | 0 | 0 | 0 | 0 | 0 | 0 | 0 | 0 | 0 | 0 | 0 | 0 | 0 | 0 | 0 | 0 |
| galR | 0 | 0 | 0 | -1 | 0 | 0 | 0 | 0 | 0 | 0 | 0 | 0 | 0 | 0 | 0 | 0 | 0 | 0 | 0 | 0 |
| hdfR | 0 | 0 | 0 | 0 | -1 | 0 | 0 | 0 | 0 | 0 | 0 | 0 | 0 | 0 | 0 | 0 | 0 | 0 | 0 | 0 |
| argP | 0 | 0 | 0 | 0 | 0 | -1 | 0 | 0 | 0 | 0 | 0 | 0 | 0 | 0 | 0 | 0 | 0 | 0 | 0 | 0 |
| basR | 0 | 0 | 0 | 0 | 0 | -1 | 0 | 0 | 0 | 0 | 0 | 0 | 0 | 0 | 0 | 0 | 0 | 0 | 0 | 0 |
| bglJ | 0 | 0 | 0 | 0 | 0 | 0 | 1 | 0 | 0 | 0 | 0 | 0 | 0 | 0 | 0 | 0 | 0 | 0 | 0 | 0 |
| pgrR | 0 | 0 | 0 | 0 | 0 | 0 | 0 | -1 | 0 | 0 | 0 | 0 | 0 | 0 | 0 | 0 | 0 | 0 | 0 | 0 |
| rcsA | 0 | 0 | 0 | 0 | 0 | 0 | 0 | 1 | 0 | 0 | 0 | 0 | 0 | 0 | 0 | 0 | 0 | 0 | 0 | 0 |
| zur | 0 | 0 | 0 | 0 | 0 | 0 | 0 | -1 | 0 | 0 | 0 | 0 | 0 | 0 | 0 | 0 | 0 | 0 | 0 | 0 |
| mlc | 0 | 0 | 0 | 0 | 0 | 0 | 0 | 1 | 0 | 0 | 0 | 0 | 0 | 0 | 0 | 0 | 0 | 0 | 0 | 0 |
| uvrY | 0 | 0 | 0 | 0 | 0 | 0 | 0 | 1 | 0 | 0 | 0 | 0 | 0 | 0 | 0 | 0 | 0 | 0 | 0 | 0 |
| ascG | 0 | 0 | 0 | 0 | 0 | 0 | 0 | 0 | -1 | 0 | 0 | 0 | 0 | 0 | 0 | 0 | 0 | 0 | 0 | 0 |
| purR | 0 | 0 | 0 | 0 | 0 | 0 | 0 | 0 | 0 | 0 | 1 | 0 | 0 | 0 | 0 | 0 | 0 | 0 | 0 | 0 |
| nadR | 0 | 0 | 0 | 0 | 0 | 0 | 0 | 0 | 0 | 0 | 0 | 0 | -1 | 0 | 0 | 0 | 0 | 0 | 0 | 0 |
| yqhC | 0 | 0 | 0 | 0 | 0 | 0 | 0 | 0 | 0 | 0 | 0 | 0 | 0 | -1 | 0 | 0 | 0 | 0 | 0 | 0 |
| creB | 0 | 0 | 0 | 0 | 0 | 0 | 0 | 0 | 0 | 0 | 0 | 0 | 0 | 0 | 0 | 0 | 1 | 0 | 0 | 0 |

**Table S3. Annotation analysis for COG identifiers.**

| Media Ref Adaptation | M9G | | | | | | | | | | Rif | | | | | | | | | |
|---|---|---|---|---|---|---|---|---|---|---|---|---|---|---|---|---|---|---|---|---|
| | pykF(C8Y) | | | | | rpoB(T1037P) | | | | | pykF(C8Y) | | | | | rpoB(T1037P) | | | | |
| COG ID | D516G | H526Y | L511R | Q148L | I572S | I572F | I572N | L511R | Q148L | S508P | D516G | H526Y | L511R | Q148L | I572S | I572F | I572N | L511R | Q148L | S508P |
| COG0076 | -1 | -1 | 1 | 1 | -1 | 1 | 1 | 1 | 0 | 0 | 0 | 0 | 1 | 1 | 0 | 1 | 0 | 1 | 1 | 0 |
| COG0589 | 1 | 0 | 0 | 0 | 1 | 1 | 1 | 0 | -1 | 1 | 0 | 1 | 0 | 0 | 0 | 0 | 0 | 0 | 0 | 0 |
| COG0175 | 0 | 0 | 0 | 0 | 0 | 0 | 0 | -1 | 0 | 0 | -1 | 0 | 0 | 0 | 0 | -1 | -1 | -1 | 0 | 0 |
| COG0600 | 0 | 0 | 0 | 0 | 0 | 0 | 0 | 0 | 0 | 0 | -1 | -1 | -1 | -1 | -1 | 0 | 0 | 0 | 0 | 0 |
| COG3677 | 1 | 1 | 0 | 0 | 0 | 0 | 0 | 0 | 0 | 1 | 0 | 1 | 0 | 0 | 0 | 0 | 1 | 0 | 0 | 0 |
| COG4575 | 0 | 0 | 1 | 0 | -1 | 0 | 0 | 0 | 0 | 0 | 0 | 0 | 0 | 0 | 0 | 1 | 0 | 1 | 1 | 0 |
| COG2963 | 1 | 0 | 0 | 0 | 0 | 0 | 0 | 0 | 0 | 0 | 0 | 0 | -1 | -1 | -1 | 0 | 0 | 0 | 0 | 0 |
| COG1177 | 0 | 0 | 0 | 0 | 0 | 0 | 0 | 0 | 0 | 0 | -1 | -1 | 0 | 0 | 0 | 0 | 0 | 0 | 0 | 0 |
| COG2207 | 0 | 0 | 0 | 0 | 0 | 0 | 1 | 0 | 0 | 1 | 0 | 0 | 0 | 0 | 0 | 0 | 0 | 0 | 0 | 0 |
| COG3039 | 0 | 0 | -1 | 0 | -1 | 0 | 0 | 0 | 0 | 0 | 0 | 0 | 0 | 0 | 0 | 0 | 0 | 0 | 0 | 0 |
| COG3098 | 0 | 0 | 0 | 0 | 0 | 0 | 0 | 0 | 0 | 0 | 0 | 1 | 0 | 0 | 0 | 0 | 0 | 0 | 0 | 1 |
| COG4525 | 0 | 0 | 0 | 0 | 0 | 0 | 0 | 0 | 0 | 0 | 0 | 0 | 0 | -1 | 0 | 0 | -1 | 0 | 0 | 0 |
| COG0002 | 0 | 0 | 0 | 0 | 1 | 0 | 0 | 0 | 0 | 0 | 0 | 0 | 0 | 0 | 0 | 0 | 0 | 0 | 0 | 0 |
| COG0028 | 0 | 0 | 0 | 0 | 0 | 0 | 0 | 0 | 0 | 0 | 0 | 0 | 0 | 1 | 0 | 0 | 0 | 0 | 0 | 0 |
| COG0031 | 0 | 0 | 0 | 0 | 0 | -1 | 0 | 0 | 0 | 0 | 0 | 0 | 0 | 0 | 0 | 0 | 0 | 0 | 0 | 0 |
| COG0183 | 0 | 0 | 0 | 0 | 0 | 0 | 0 | 0 | 1 | 0 | 0 | 0 | 0 | 0 | 0 | 0 | 0 | 0 | 0 | 0 |
| COG0246 | 0 | 1 | 0 | 0 | 0 | 0 | 0 | 0 | 0 | 0 | 0 | 0 | 0 | 0 | 0 | 0 | 0 | 0 | 0 | 0 |
| COG0316 | 0 | 0 | 0 | 0 | 0 | 0 | 0 | 0 | 0 | 0 | -1 | 0 | 0 | 0 | 0 | 0 | 0 | 0 | 0 | 0 |
| COG0398 | 0 | 0 | 0 | -1 | 0 | 0 | 0 | 0 | 0 | 0 | 0 | 0 | 0 | 0 | 0 | 0 | 0 | 0 | 0 | 0 |
| COG0502 | 0 | 0 | 0 | 0 | -1 | 0 | 0 | 0 | 0 | 0 | 0 | 0 | 0 | 0 | 0 | 0 | 0 | 0 | 0 | 0 |
| COG0508 | 0 | 0 | 0 | 0 | -1 | 0 | 0 | 0 | 0 | 0 | 0 | 0 | 0 | 0 | 0 | 0 | 0 | 0 | 0 | 0 |
| COG0516 | 0 | 0 | 0 | 0 | 0 | 0 | 0 | 0 | 0 | 0 | 0 | -1 | 0 | 0 | 0 | 0 | 0 | 0 | 0 | 0 |
| COG0527 | 0 | 0 | 0 | 0 | 0 | -1 | 0 | 0 | 0 | 0 | 0 | 0 | 0 | 0 | 0 | 0 | 0 | 0 | 0 | 0 |
| COG0583 | 0 | 0 | 0 | 0 | 0 | 0 | 1 | 0 | 0 | 0 | 0 | 0 | 0 | 0 | 0 | 0 | 0 | 0 | 0 | 0 |
| COG0687 | 0 | 0 | 0 | 1 | 0 | 0 | 0 | 0 | 0 | 0 | 0 | 0 | 0 | 0 | 0 | 0 | 0 | 0 | 0 | 0 |
| COG0719 | 0 | 0 | 0 | 0 | -1 | 0 | 0 | 0 | 0 | 0 | 0 | 0 | 0 | 0 | 0 | 0 | 0 | 0 | 0 | 0 |
| COG0782 | 0 | 0 | -1 | 0 | 0 | 0 | 0 | 0 | 0 | 0 | 0 | 0 | 0 | 0 | 0 | 0 | 0 | 0 | 0 | 0 |
| COG0859 | 0 | 0 | -1 | 0 | 0 | 0 | 0 | 0 | 0 | 0 | 0 | 0 | 0 | 0 | 0 | 0 | 0 | 0 | 0 | 0 |
| COG0861 | 0 | 0 | 0 | 0 | 0 | 0 | 0 | -1 | 0 | 0 | 0 | 0 | 0 | 0 | 0 | 0 | 0 | 0 | 0 | 0 |
| COG1012 | 0 | 0 | 0 | 0 | 0 | 0 | 0 | 0 | 0 | 0 | 0 | -1 | 0 | 0 | 0 | 0 | 0 | 0 | 0 | 0 |
| COG1253 | 0 | 0 | 0 | 0 | 0 | 0 | 0 | -1 | 0 | 0 | 0 | 0 | 0 | 0 | 0 | 0 | 0 | 0 | 0 | 0 |
| COG1442 | 0 | 0 | -1 | 0 | 0 | 0 | 0 | 0 | 0 | 0 | 0 | 0 | 0 | 0 | 0 | 0 | 0 | 0 | 0 | 0 |
| COG2391 | 0 | 0 | 0 | 0 | 0 | 0 | 0 | 0 | 0 | 0 | 0 | 0 | 0 | 0 | 0 | 0 | 0 | -1 | 0 | 0 |
| COG2721 | 1 | 0 | 0 | 0 | 0 | 0 | 0 | 0 | 0 | 0 | 0 | 0 | 0 | 0 | 0 | 0 | 0 | 0 | 0 | 0 |
| COG3038 | 0 | 0 | 1 | 0 | 0 | 0 | 0 | 0 | 0 | 0 | 0 | 0 | 0 | 0 | 0 | 0 | 0 | 0 | 0 | 0 |
| COG3685 | 0 | 0 | 0 | 0 | 0 | 0 | 0 | 0 | 0 | 0 | 0 | 0 | 0 | 0 | -1 | 0 | 0 | 0 | 0 | 0 |
| COG4531 | 1 | 0 | 0 | 0 | 0 | 0 | 0 | 0 | 0 | 0 | 0 | 0 | 0 | 0 | 0 | 0 | 0 | 0 | 0 | 0 |
| COG4580 | 0 | 0 | 0 | 0 | 0 | 0 | 0 | 0 | 0 | 0 | 0 | 0 | 0 | 0 | 0 | 0 | 0 | 0 | 0 | -1 |

**Table S4. Annotation analysis for COG Groups.**

| Media | M9G | | | | | | | | | | Rif | | | | | | | | | |
|---|---|---|---|---|---|---|---|---|---|---|---|---|---|---|---|---|---|---|---|---|
| Ref | pykF(C8Y) | | | | | rpoB(T1037P) | | | | | pykF(C8Y) | | | | | rpoB(T1037P) | | | | |
| Adaptation | D516G | H526Y | L511R | Q148L | I572S | I572F | I572N | L511R | Q148L | S508P | D516G | H526Y | L511R | Q148L | I572S | I572F | I572N | L511R | Q148L | S508P |
| COG Group | | | | | | | | | | | | | | | | | | | | |
| J | -1 | -1 | -1 | 1 | -1 | -1 | -1 | -1 | 0 | -1 | -1 | -1 | 1 | 1 | 1 | 1 | 0 | 1 | 1 | 1 |
| F | -1 | -1 | 0 | 1 | 0 | 1 | -1 | 0 | 1 | -1 | -1 | -1 | 0 | 1 | 0 | 0 | 0 | 0 | 1 | 0 |
| E | -1 | 0 | 0 | 1 | 0 | -1 | -1 | 0 | 0 | -1 | -1 | -1 | 0 | 1 | 0 | 0 | -1 | 0 | 1 | 0 |
| H | -1 | -1 | -1 | 1 | -1 | 0 | -1 | 0 | 0 | 0 | -1 | 0 | 0 | 0 | 0 | 0 | 0 | 0 | 1 | 1 |
| K | 1 | 0 | -1 | 0 | 0 | 1 | 1 | 0 | 0 | 1 | 0 | 0 | -1 | -1 | 0 | 0 | 1 | 0 | 0 | 1 |
| L | -1 | 0 | -1 | -1 | 0 | 0 | 0 | -1 | -1 | 0 | 0 | 1 | 0 | 0 | 0 | 0 | 1 | 0 | 0 | 0 |
| T | 0 | 0 | -1 | 0 | 0 | 1 | 1 | 1 | 0 | 1 | 0 | 0 | 0 | 0 | 0 | 0 | 1 | 0 | 0 | 1 |
| M | -1 | 0 | -1 | 0 | 0 | -1 | 0 | 0 | 0 | -1 | 0 | 0 | 0 | 1 | 0 | 0 | 0 | 0 | 1 | 0 |
| D | 0 | 0 | -1 | 0 | -1 | -1 | 0 | 0 | -1 | 0 | 0 | 0 | 0 | 0 | 0 | 0 | 1 | 0 | 0 | 0 |
| Q | 0 | 0 | 0 | 0 | 0 | 0 | 0 | 0 | 0 | 0 | -1 | 0 | 0 | 0 | -1 | 0 | 0 | 1 | 0 | 1 |
| O | 0 | 0 | -1 | 0 | -1 | 0 | 0 | 0 | 0 | 0 | -1 | 0 | 0 | 0 | 0 | 0 | 0 | 0 | 0 | 0 |
| R | 0 | 0 | 0 | 0 | 0 | 0 | 1 | 0 | 0 | 0 | 0 | 0 | -1 | -1 | 0 | 0 | 0 | 0 | 0 | 0 |
| I | 0 | 0 | 0 | 1 | 0 | -1 | 0 | 0 | 0 | 0 | 0 | 0 | 0 | 0 | 0 | 0 | 0 | 0 | 0 | 0 |
| P | 0 | 0 | 0 | 0 | 0 | 0 | 0 | -1 | 0 | 0 | 0 | 0 | 0 | -1 | 0 | 0 | 0 | 0 | 0 | 0 |
| U | 0 | 0 | 0 | 0 | 0 | -1 | 0 | -1 | 0 | 0 | 0 | 0 | 0 | 0 | 0 | 0 | 0 | 0 | 0 | 0 |
| V | 0 | 1 | 0 | 0 | 0 | 0 | 0 | 0 | 0 | 0 | 0 | 0 | -1 | 0 | 0 | 0 | 0 | 0 | 0 | 0 |
| C | 0 | 0 | 0 | 0 | 0 | 0 | 0 | 0 | 0 | 0 | 0 | -1 | 0 | 0 | 0 | 0 | 0 | 0 | 0 | 0 |

**Table S5. Annotation analysis for PANTHER pathways.**

| Media | M9G | | | | | | | | | | Rif | | | | | | | | | |
|---|---|---|---|---|---|---|---|---|---|---|---|---|---|---|---|---|---|---|---|---|
| Ref | pykF(C8Y) | | | | | rpoB(T1037P) | | | | | pykF(C8Y) | | | | | rpoB(T1037P) | | | | |
| Adaptation | D516G | H526Y | L511R | Q148L | I572S | I572F | I572N | L511R | Q148L | S508P | D516G | H526Y | L511R | Q148L | I572S | I572F | I572N | L511R | Q148L | S508P |
| Pathway ID | | | | | | | | | | | | | | | | | | | | |
| P02747 | 1 | 1 | 0 | -1 | 1 | 1 | 0 | 0 | -1 | 0 | 1 | 1 | 1 | 1 | 1 | 1 | 1 | 1 | 0 | 1 |
| P02738 | -1 | -1 | 0 | 1 | 0 | 1 | -1 | 0 | 1 | -1 | 0 | 0 | 0 | 1 | 0 | 0 | 0 | 1 | 1 | 0 |
| P04372 | 0 | 0 | 0 | 1 | 0 | 0 | 0 | 0 | 0 | -1 | -1 | -1 | 0 | 1 | 0 | 1 | 0 | 1 | 1 | 1 |
| P02757 | 0 | 0 | 0 | 1 | 0 | 0 | -1 | 0 | 0 | -1 | 0 | 0 | 0 | 1 | 0 | 1 | 0 | 0 | 1 | 0 |
| P02728 | 0 | 0 | 0 | 0 | 0 | 0 | 0 | 0 | 0 | -1 | 0 | -1 | 0 | 1 | 0 | 0 | -1 | -1 | 0 | -1 |
| P02748 | 0 | 0 | 0 | 0 | 0 | -1 | -1 | 0 | 0 | 0 | -1 | 0 | 0 | 1 | 0 | 0 | 0 | 0 | 1 | 0 |
| P02749 | 0 | 0 | 0 | 0 | -1 | -1 | -1 | 0 | 0 | 0 | 0 | 0 | 0 | 0 | 0 | 0 | 0 | 1 | 1 | 0 |
| P02726 | 0 | 0 | 0 | 1 | -1 | 0 | 0 | 0 | 0 | 0 | -1 | -1 | 0 | 0 | 0 | 0 | 0 | 0 | 1 | 0 |
| P02785 | 0 | 0 | 0 | 0 | 0 | -1 | -1 | 0 | 0 | 0 | -1 | 0 | 0 | 0 | 0 | 0 | 0 | 0 | 1 | 0 |
| P02776 | 0 | 0 | 0 | 0 | 0 | 0 | -1 | 0 | 0 | -1 | 0 | 0 | 0 | 1 | 0 | 0 | 0 | 0 | 1 | 0 |
| P00024 | 0 | 0 | 0 | -1 | 0 | 0 | 0 | 0 | 0 | 0 | 1 | 1 | 0 | 0 | 1 | 0 | 0 | 0 | 0 | 0 |
| P02740 | 0 | 0 | 0 | 0 | 0 | 0 | 0 | 0 | 0 | 0 | 0 | 0 | 1 | 1 | 0 | 0 | 0 | 0 | 1 | 0 |
| P02739 | 0 | 0 | 0 | 0 | 0 | 0 | 0 | 0 | 0 | 0 | 0 | 0 | 1 | 1 | 0 | 0 | 0 | 1 | 1 | 0 |
| P02731 | 0 | 0 | 0 | 0 | -1 | 0 | 0 | 0 | 0 | 0 | -1 | 0 | 0 | 0 | 0 | 0 | 0 | 0 | 1 | 0 |
| P02721 | 0 | 0 | 0 | 0 | 0 | 0 | -1 | 0 | 0 | 0 | 0 | 0 | 0 | 1 | 0 | 0 | 0 | 0 | 1 | 0 |
| P02781 | 0 | 0 | 0 | 0 | 0 | -1 | -1 | 0 | 0 | 0 | 0 | 0 | 0 | 0 | 0 | 0 | 0 | 0 | 0 | 0 |
| P02778 | 0 | 0 | 0 | 0 | 0 | 0 | 0 | 0 | 0 | 0 | 0 | 0 | 0 | 0 | 0 | 0 | 0 | -1 | 0 | 0 |
| P02725 | 0 | 0 | 1 | 0 | 0 | 0 | 0 | 0 | 0 | 0 | 0 | 0 | 0 | 0 | 0 | 0 | 0 | 0 | 0 | 0 |
| P02763 | 0 | 0 | 0 | 0 | -1 | 0 | 0 | 0 | 0 | 0 | 0 | 0 | 0 | 0 | 0 | 0 | 0 | 0 | 0 | 0 |
| P02783 | 0 | 0 | 0 | 0 | 0 | 0 | -1 | 0 | 0 | 0 | 0 | 0 | 0 | 0 | 0 | 0 | 0 | 0 | 0 | 0 |
| P02734 | 0 | 0 | 0 | 0 | 0 | 0 | 0 | 0 | 0 | -1 | 0 | 0 | 0 | 0 | 0 | 0 | 0 | 0 | 0 | 0 |
| P02751 | 0 | 0 | 0 | 0 | 0 | 0 | 0 | 0 | 0 | 0 | 0 | 0 | 0 | 1 | 0 | 0 | 0 | 0 | 0 | 0 |

**Table S6. Annotation analysis for PANTHER protein classes.**

| Media | M9G | | | | | | | | | | Rif | | | | | | | | | |
|---|---|---|---|---|---|---|---|---|---|---|---|---|---|---|---|---|---|---|---|---|
| Ref | D516G | pykF(C8Y) | | | | rpoB(T1037P) | | | | | D516G | pykF(C8Y) | | | | rpoB(T1037P) | | | | |
| Adaptation | | H526Y | L511R | Q148L | I572S | I572F | I572N | L511R | Q148L | S508P | | H526Y | L511R | Q148L | I572S | I572F | I572N | L511R | Q148L | S508P |
| **Protein Class** | | | | | | | | | | | | | | | | | | | | |
| PC00202 | -1 | -1 | 0 | 1 | -1 | -1 | -1 | -1 | 0 | -1 | -1 | -1 | 1 | 1 | 0 | 1 | -1 | 1 | 1 | 0 |
| PC00031 | -1 | -1 | 0 | 1 | -1 | -1 | -1 | -1 | 0 | -1 | -1 | -1 | 1 | 1 | 0 | 1 | 0 | 1 | 1 | 1 |
| PC00171 | -1 | -1 | 0 | 1 | -1 | -1 | -1 | -1 | 0 | -1 | -1 | 0 | 1 | 1 | 0 | 1 | 0 | 1 | 1 | 1 |
| PC00211 | -1 | -1 | 1 | 0 | 0 | -1 | 0 | 0 | 0 | -1 | 0 | 0 | 1 | 0 | 0 | -1 | 0 | 0 | 0 | -1 |
| PC00220 | 0 | 0 | 0 | 1 | 0 | 0 | 0 | 0 | 0 | -1 | -1 | 0 | 0 | 1 | 0 | 0 | 0 | 1 | 1 | 1 |
| PC00095 | 0 | 0 | -1 | 0 | 0 | -1 | -1 | 0 | 0 | 0 | 0 | 0 | 0 | 1 | 0 | 0 | 0 | 0 | 1 | 1 |
| PC00121 | 0 | 0 | 0 | 0 | 0 | 0 | -1 | 0 | 0 | 0 | -1 | 0 | 0 | 1 | 0 | 0 | 0 | 0 | 0 | 0 |
| PC00223 | -1 | 0 | 0 | 0 | 0 | -1 | -1 | 0 | 0 | -1 | 0 | 0 | 0 | 1 | 0 | 0 | 0 | 0 | 0 | 0 |
| PC00142 | 0 | 0 | 0 | 1 | 0 | 0 | -1 | 0 | 0 | -1 | -1 | 0 | 0 | 1 | 0 | 0 | 0 | 0 | 1 | 0 |
| PC00222 | 0 | 0 | 0 | 0 | 0 | -1 | -1 | -1 | 0 | 0 | 0 | 0 | 0 | 1 | 0 | 0 | 0 | 0 | 0 | 0 |
| PC00020 | 0 | 0 | -1 | 0 | 0 | -1 | -1 | -1 | 0 | 0 | 0 | 0 | 0 | 0 | 0 | 0 | 0 | 0 | 0 | 0 |
| PC00116 | 0 | 0 | 0 | 0 | 0 | 0 | 0 | 0 | 0 | 1 | 0 | 0 | -1 | -1 | 0 | 0 | 0 | 0 | 0 | 0 |
| PC00218 | 0 | 0 | 0 | 0 | 0 | 0 | 1 | 0 | 0 | 1 | 0 | 0 | -1 | -1 | 0 | 0 | 0 | 0 | 0 | 0 |
| PC00175 | 0 | 0 | 0 | 1 | 0 | 0 | 0 | 0 | 0 | 0 | -1 | 0 | 0 | 0 | 0 | 0 | 0 | 0 | 0 | 0 |
| PC00144 | 0 | 0 | 0 | 1 | 0 | 0 | 0 | 0 | 0 | 0 | -1 | 0 | 0 | 0 | 0 | 0 | 0 | 0 | 0 | 0 |
| PC00176 | 0 | 0 | 1 | 1 | 0 | 0 | 0 | 0 | 0 | 0 | -1 | 0 | 0 | 1 | 0 | 0 | 0 | 0 | 0 | 0 |
| PC00092 | 0 | 0 | 1 | 1 | 0 | 0 | 0 | 0 | 0 | 0 | -1 | -1 | 0 | 0 | 0 | 0 | 0 | 0 | 0 | 0 |
| PC00113 | 0 | 0 | 0 | 0 | 0 | 0 | 0 | 0 | 0 | 0 | 0 | 0 | 0 | 1 | 0 | 1 | 0 | 0 | 1 | 0 |
| PC00022 | 0 | 0 | 0 | 0 | 0 | 0 | 0 | 0 | 0 | 0 | 0 | 0 | 0 | 1 | 0 | 1 | 0 | 0 | 1 | 0 |
| PC00002 | 0 | 0 | 0 | 0 | 0 | 0 | -1 | 0 | 0 | 0 | 0 | 0 | 0 | 0 | 0 | 0 | 0 | 0 | 0 | 0 |
| PC00190 | 0 | 0 | 0 | 0 | 0 | 0 | -1 | 0 | 0 | 0 | 0 | 0 | 0 | 0 | 0 | 0 | 0 | 0 | 0 | 0 |
| PC00067 | 0 | 0 | 0 | 0 | 1 | 0 | 0 | 0 | 1 | 0 | 0 | 0 | 0 | 0 | 0 | 0 | 0 | 0 | 0 | 0 |
| PC00088 | 0 | 0 | 0 | 0 | 0 | 0 | -1 | 0 | 0 | 0 | 0 | 0 | 0 | 0 | 0 | 0 | 0 | 0 | 0 | 0 |
| PC00042 | 0 | 0 | 0 | 0 | 0 | 0 | -1 | 0 | 0 | -1 | 0 | 0 | 0 | 1 | 0 | 0 | 0 | 0 | 0 | 0 |
| PC00046 | 0 | 0 | 0 | 0 | 0 | 0 | 0 | 0 | 0 | -1 | 0 | -1 | 0 | 0 | 0 | 0 | 0 | 0 | 0 | 0 |
| PC00246 | 0 | 0 | 0 | 0 | 0 | 0 | 0 | 0 | 0 | 0 | 0 | 0 | 0 | 0 | 0 | 0 | 0 | 0 | 0 | 0 |
| PC00174 | 0 | 0 | -1 | 0 | 0 | 0 | 0 | 0 | 0 | 0 | 0 | 0 | 0 | 0 | 0 | 0 | 0 | 0 | 0 | 0 |
| PC00155 | 0 | 0 | 0 | 1 | 0 | 0 | 0 | 0 | 0 | 0 | 0 | 0 | 0 | 0 | 0 | 0 | 0 | 0 | 0 | 0 |
| PC00048 | 0 | 0 | 0 | 0 | 0 | 0 | 0 | 0 | 0 | 0 | 0 | 0 | 0 | 0 | 0 | 0 | 0 | 0 | 0 | 0 |
| PC00047 | 0 | 0 | 0 | 0 | 0 | 0 | -1 | 0 | 0 | 0 | 0 | 0 | 0 | 0 | 0 | 0 | 0 | 0 | 0 | 0 |
| PC00203 | 0 | 0 | 0 | 0 | 0 | 0 | -1 | 0 | 0 | 0 | 0 | 0 | 0 | 0 | 0 | 0 | 0 | 0 | 0 | 0 |
| PC00224 | 0 | 0 | 0 | 0 | 0 | 0 | -1 | 0 | 0 | 0 | 0 | 0 | 0 | 1 | 0 | 0 | 0 | 0 | 0 | 0 |
| PC00096 | 0 | 0 | 0 | 0 | 0 | 0 | 0 | 0 | 0 | 0 | 0 | 0 | 0 | 0 | 0 | 0 | 0 | 0 | 0 | 0 |
| PC00003 | 0 | 0 | 0 | 0 | 0 | 0 | 0 | 0 | 0 | 0 | -1 | 0 | 0 | 0 | 0 | 0 | 0 | 0 | 0 | 0 |
| PC00227 | 0 | 0 | 0 | 0 | 0 | 0 | 0 | 0 | 0 | 0 | 0 | -1 | 0 | 0 | 0 | 0 | 0 | 0 | 0 | 0 |
| PC00091 | 0 | 0 | 0 | 0 | 0 | 0 | 0 | 0 | 0 | 0 | 0 | 0 | 0 | 0 | 0 | 0 | -1 | 0 | 0 | 0 |
| PC00198 | 0 | 0 | 0 | 0 | 0 | 0 | 0 | 0 | 0 | 0 | 0 | 0 | 0 | 0 | 0 | 0 | -1 | 0 | 0 | 0 |
| PC00081 | 0 | 0 | 0 | 0 | 0 | 0 | 0 | 0 | 0 | 0 | 0 | 0 | 0 | 1 | 0 | 0 | 0 | 0 | 1 | 0 |
| PC00216 | 0 | 0 | 0 | 1 | 0 | 0 | 0 | 0 | 0 | 0 | -1 | 0 | 0 | 0 | 0 | 0 | 0 | 0 | 0 | 0 |
| PC00208 | 0 | 0 | -1 | -1 | 0 | 0 | 0 | 0 | 0 | 0 | 0 | 0 | 0 | 0 | 0 | 0 | 0 | 0 | 0 | 0 |
| PC00135 | 0 | 0 | 0 | 0 | 0 | 0 | 0 | 0 | 0 | 0 | 0 | 0 | 0 | 1 | 0 | 0 | 0 | 0 | 0 | 0 |
| PC00068 | 0 | 0 | 0 | 0 | 0 | 0 | 0 | 0 | 0 | 0 | 0 | 0 | 0 | 1 | 0 | 0 | 0 | 0 | 0 | 0 |

**Table S7. Annotation analysis for origons, further illustrated in Fig. 4.**

| Media | M9G | | | | | | | | | | Rif | | | | | | | | | |
|---|---|---|---|---|---|---|---|---|---|---|---|---|---|---|---|---|---|---|---|---|
| Ref | pykF(C8Y) | | | | | rpoB(T1037P) | | | | | pykF(C8Y) | | | | | rpoB(T1037P) | | | | |
| Adaptation | D516G | H526Y | I572S | L511R | Q148L | I572F | I572N | L511R | Q148L | S508P | D516G | H526Y | I572S | L511R | Q148L | I572F | I572N | L511R | Q148L | S508P |
| **Regulator** | | | | | | | | | | | | | | | | | | | | |
| phoP | 0 | 0 | -1 | 0 | 0 | 0 | 0 | -1 | 0 | 0 | -1 | 0 | -1 | -1 | -1 | -1 | -1 | -1 | -1 | -1 |
| acrR | 1 | 1 | 0 | 0 | -1 | 1 | 1 | 0 | 0 | 1 | 1 | 1 | 0 | 0 | 0 | 0 | 0 | 0 | -1 | 0 |
| phoB | 0 | -1 | 0 | 0 | -1 | 0 | 0 | 0 | -1 | 1 | 1 | 1 | 1 | 0 | 0 | 0 | 1 | 0 | 0 | 0 |
| argR | 0 | 0 | -1 | 0 | 0 | 1 | 1 | 0 | 0 | 1 | 0 | 0 | 0 | 0 | 0 | 0 | 0 | 1 | 0 | 1 |
| iscR | 0 | 0 | -1 | 0 | 0 | 0 | 0 | 0 | 0 | 0 | 0 | 0 | 0 | 0 | 0 | 1 | 1 | 1 | 1 | 1 |
| csiR | 0 | 0 | 0 | 0 | -1 | 0 | 0 | 0 | 0 | 0 | 0 | 0 | 0 | 0 | 0 | -1 | 0 | -1 | -1 | -1 |
| lrhA | 0 | 1 | 0 | 0 | 1 | 0 | 0 | 0 | 0 | 1 | -1 | 0 | 0 | 0 | 0 | 0 | 0 | 0 | 1 | 0 |
| modE | -1 | 0 | 0 | 0 | 1 | 0 | 0 | 0 | 0 | 0 | -1 | -1 | 0 | 0 | 0 | 0 | -1 | 0 | 0 | 0 |
| nsrR | 0 | 0 | 0 | 0 | 1 | 0 | 0 | 0 | 0 | -1 | -1 | -1 | 0 | 0 | 0 | 0 | 0 | 0 | 1 | 0 |
| torR | 1 | 1 | 0 | 0 | -1 | 0 | 0 | -1 | 0 | 1 | 0 | 0 | 0 | 0 | 0 | 0 | 0 | 0 | 0 | 0 |
| birA | 1 | 0 | 1 | 0 | 0 | 0 | 0 | 0 | 0 | 0 | 1 | 0 | 0 | 0 | 0 | 0 | 0 | 0 | -1 | 0 |
| lexA | 0 | 0 | 0 | 1 | 1 | 0 | 0 | 0 | 0 | 0 | 0 | -1 | 0 | 1 | 0 | 0 | 0 | 0 | 0 | 0 |
| slyA | 1 | 1 | 0 | 0 | 0 | 0 | 0 | 0 | 0 | 1 | 0 | 0 | 0 | 0 | 0 | 0 | 0 | 0 | 0 | 1 |
| zur | -1 | -1 | 0 | 0 | 0 | 0 | 0 | -1 | 0 | -1 | 0 | 0 | 0 | 0 | 0 | 0 | 0 | 0 | 0 | 0 |
| rutR | 0 | 1 | 0 | 0 | 0 | 0 | 0 | -1 | 0 | 1 | 0 | 0 | 0 | 0 | 0 | 0 | 0 | 0 | 0 | 0 |
| sdiA | 1 | 1 | 0 | 0 | 0 | 0 | 0 | -1 | 0 | 0 | 0 | 0 | 0 | 0 | 0 | 0 | 0 | 0 | 0 | 0 |
| cpxR | 0 | 0 | 0 | 0 | 0 | 0 | 0 | 0 | 0 | 0 | 0 | 0 | 0 | -1 | 0 | 0 | 0 | 0 | 0 | 1 |
| fadR | 0 | 0 | 0 | 0 | -1 | 0 | -1 | 0 | 0 | 0 | 0 | 0 | 0 | 0 | 0 | 0 | 0 | 0 | 0 | 0 |
| nadR | 0 | 0 | 0 | 0 | -1 | 0 | 0 | 0 | 0 | 0 | 0 | 0 | 0 | -1 | 0 | 0 | 0 | 0 | 0 | 0 |
| trpR | 0 | 0 | 0 | 0 | 0 | 0 | 0 | -1 | -1 | 0 | 0 | 0 | 0 | 0 | 0 | 0 | 0 | 0 | 0 | 0 |
| yedW | 0 | 0 | 0 | 0 | 0 | 0 | 0 | 0 | -1 | 0 | 0 | 0 | 0 | 0 | 0 | -1 | 0 | 0 | 0 | 0 |
| basR | 0 | 0 | 1 | 0 | 0 | 0 | 0 | 0 | 0 | 0 | 0 | 0 | 0 | 0 | 0 | 0 | 0 | 0 | 0 | 0 |
| mcbR | 0 | 0 | 0 | 0 | 0 | 0 | 0 | 0 | 0 | 0 | 0 | -1 | 0 | 0 | 0 | 0 | 0 | 0 | 0 | 0 |
| rcdA | 0 | 0 | 1 | 0 | 0 | 0 | 0 | 0 | 0 | 0 | 0 | 0 | 0 | 0 | 0 | 0 | 0 | 0 | 0 | 0 |
| tdcR | 0 | 0 | 0 | 0 | 0 | 0 | 0 | 0 | 1 | 0 | 0 | 0 | 0 | 0 | 0 | 0 | 0 | 0 | 0 | 0 |
| ydcN | 0 | 0 | 1 | 0 | 0 | 0 | 0 | 0 | 0 | 0 | 0 | 0 | 0 | 0 | 0 | 0 | 0 | 0 | 0 | 0 |

**Table S8. Annotation analysis for modulons.**

| Media | M9G | | | | | | | | | | Rif | | | | | | | | | |
|---|---|---|---|---|---|---|---|---|---|---|---|---|---|---|---|---|---|---|---|---|
| Ref | pykF(C8Y) | | | | | rpoB(T1037P) | | | | | pykF(C8Y) | | | | | rpoB(T1037P) | | | | |
| Adaptation | D516G | H526Y | I572S | L511R | Q148L | I572F | I572N | L511R | Q148L | S508P | D516G | H526Y | I572S | L511R | Q148L | I572F | I572N | L511R | Q148L | S508P |
| **Modulon** | | | | | | | | | | | | | | | | | | | | |
| RpoS | -1 | -1 | -1 | 1 | 1 | 1 | 1 | 1 | 0 | -1 | -1 | -1 | 0 | 1 | 1 | 1 | 1 | 1 | 1 | 1 |
| GadEWX | -1 | -1 | -1 | 1 | 1 | 1 | 1 | 1 | 0 | 0 | 1 | 0 | 1 | 1 | 1 | 1 | 1 | 1 | 1 | 1 |
| translation | -1 | -1 | -1 | 0 | 1 | -1 | -1 | -1 | 0 | -1 | -1 | -1 | 0 | 1 | 1 | -1 | -1 | 1 | 1 | 0 |
| CysB | 0 | 1 | -1 | 0 | 0 | -1 | -1 | -1 | 0 | 0 | -1 | -1 | -1 | -1 | -1 | -1 | -1 | -1 | -1 | 0 |
| GadWX | -1 | -1 | 0 | 1 | 0 | 1 | 1 | 1 | 0 | 0 | 1 | 1 | 1 | 1 | 1 | 1 | 1 | 1 | 0 | 1 |
| iron-related | -1 | 0 | 0 | 1 | 0 | -1 | -1 | 0 | -1 | -1 | -1 | -1 | -1 | -1 | -1 | -1 | -1 | -1 | 0 | -1 |
| Fur-1 | -1 | -1 | -1 | 0 | 0 | 0 | -1 | 0 | 0 | -1 | -1 | -1 | 0 | 1 | 1 | 1 | 1 | 1 | 1 | 1 |
| uncharacterized-5 | -1 | -1 | 0 | 1 | 0 | 0 | -1 | -1 | 1 | -1 | 0 | -1 | 1 | 1 | 1 | 1 | 1 | 0 | 0 | -1 |
| PurR-2 | -1 | -1 | -1 | 1 | 1 | 1 | 0 | 0 | 0 | -1 | -1 | -1 | 0 | 1 | 1 | 0 | 0 | 1 | 1 | 1 |
| ArcA-1 | -1 | 0 | 0 | 1 | 1 | -1 | -1 | 0 | 1 | -1 | -1 | -1 | 0 | 0 | 1 | 1 | 1 | 1 | 1 | 1 |
| NarL | 1 | 0 | 1 | 0 | -1 | 1 | 0 | 0 | -1 | 1 | 1 | 1 | 0 | 0 | 0 | -1 | -1 | -1 | -1 | -1 |
| PurR-1 | -1 | -1 | 0 | 1 | 1 | 0 | 0 | 0 | 0 | -1 | -1 | -1 | 0 | 0 | 1 | 0 | -1 | 0 | 1 | 1 |
| His-tRNA | 1 | 1 | 1 | -1 | -1 | 1 | 0 | 0 | 0 | 0 | 1 | 1 | 1 | 0 | 0 | 1 | 1 | 0 | 0 | 1 |
| Cbl+CysB | 1 | 1 | 0 | 0 | 0 | 0 | 0 | 0 | 0 | 0 | -1 | -1 | -1 | -1 | -1 | -1 | -1 | -1 | -1 | -1 |
| FlhDC | -1 | -1 | 0 | 1 | 0 | -1 | 0 | 0 | 0 | -1 | -1 | -1 | 0 | 1 | 0 | -1 | -1 | 0 | 0 | 1 |
| efeU-repair | -1 | -1 | -1 | 1 | 1 | 0 | -1 | 0 | 0 | 0 | -1 | -1 | -1 | 0 | 0 | 0 | 1 | 1 | 1 | 0 |
| Leu/Ile | -1 | 0 | -1 | 0 | 1 | -1 | -1 | 0 | 0 | -1 | -1 | 0 | 0 | 0 | 1 | 0 | -1 | 1 | 1 | 1 |
| proVWX | 1 | 0 | 0 | -1 | -1 | 1 | 1 | 0 | 0 | 0 | 0 | 0 | 1 | -1 | 0 | 1 | 1 | 1 | 0 | 1 |
| RpoH | -1 | -1 | -1 | 1 | 0 | 0 | -1 | 0 | 1 | 0 | -1 | -1 | 0 | 0 | 0 | 1 | -1 | 0 | 1 | 0 |
| membrane | 0 | 0 | 0 | 0 | -1 | 1 | 0 | 0 | 0 | 0 | 1 | 1 | 1 | 1 | 0 | 1 | 1 | 1 | 1 | 1 |
| fimbriae | 0 | -1 | 1 | 0 | 0 | 0 | 1 | 0 | 0 | 0 | 1 | 1 | 1 | 1 | 0 | 1 | 1 | 1 | 0 | 1 |
| uncharacterized-3 | 0 | 0 | 0 | -1 | -1 | -1 | -1 | 0 | 0 | 0 | 1 | 1 | 0 | 0 | -1 | 0 | 0 | -1 | -1 | -1 |
| MalT | 0 | 0 | 0 | 0 | 0 | 0 | -1 | 0 | 0 | -1 | 1 | 1 | 1 | 1 | 0 | 0 | -1 | -1 | -1 | -1 |
| ArgR | -1 | -1 | 1 | 0 | 0 | 0 | 0 | 0 | 0 | 0 | 0 | -1 | 0 | 0 | 1 | 0 | 0 | -1 | -1 | -1 |
| Crp-1 | 0 | 0 | 0 | 1 | 1 | 0 | -1 | 0 | 0 | -1 | 0 | 0 | 0 | 1 | 1 | -1 | -1 | 0 | 0 | -1 |
| Fnr | 0 | -1 | 1 | 0 | -1 | 1 | 0 | 0 | -1 | 0 | 1 | 1 | 0 | 0 | 0 | 0 | 0 | -1 | -1 | 0 |
| uncharacterized-6 | 1 | 0 | 0 | 0 | 0 | 1 | 1 | 0 | 0 | 0 | 0 | 0 | 0 | -1 | -1 | -1 | 0 | 0 | -1 | 0 |
| PuuR | 0 | 0 | 0 | 0 | 1 | -1 | 0 | 0 | 1 | 0 | -1 | -1 | 0 | 0 | 0 | 0 | 0 | 1 | 1 | 1 |
| ArcA-2 | 0 | 0 | 1 | 0 | -1 | 0 | 1 | 0 | 0 | 0 | 1 | 1 | 0 | 0 | 0 | 0 | 0 | -1 | -1 | -1 |
| Lrp | 0 | 0 | 1 | 0 | -1 | 1 | 0 | 0 | 0 | 0 | 1 | 1 | 0 | 0 | 0 | 0 | 0 | -1 | -1 | -1 |
| CecR | 0 | 0 | 0 | -1 | 0 | 1 | 1 | 0 | 0 | 0 | 0 | 0 | 0 | 0 | 0 | 1 | 1 | 1 | 0 | 0 |
| OxyR | 1 | 1 | 0 | 0 | 0 | 0 | 1 | 0 | 0 | 1 | 0 | 1 | 0 | 0 | -1 | 0 | 0 | 0 | 0 | 0 |
| Zinc | 1 | 1 | 0 | 1 | 0 | 0 | 1 | 1 | 0 | 0 | 0 | 1 | 0 | 0 | 0 | 0 | 0 | 0 | 0 | 0 |
| curli | 0 | 1 | -1 | 0 | -1 | -1 | 0 | 0 | 0 | 1 | 0 | 0 | 0 | 0 | 0 | -1 | -1 | 0 | 0 | 0 |
| NikR | -1 | -1 | 0 | 0 | -1 | 0 | 0 | 0 | 0 | 0 | 1 | 1 | 0 | 0 | 0 | 0 | 0 | -1 | -1 | 0 |
| Thiamine | 0 | 0 | -1 | 0 | 1 | 0 | 0 | 0 | 0 | 0 | -1 | -1 | 0 | 0 | 0 | 1 | 0 | 1 | 0 | 1 |
| Fur-2 | 0 | 0 | 0 | -1 | 0 | 0 | 0 | 0 | 0 | 0 | 1 | 1 | 0 | 0 | 0 | 1 | 1 | 1 | 0 | 1 |
| Tryptophan | 0 | 0 | 0 | 0 | 1 | 0 | -1 | 1 | 0 | -1 | -1 | 0 | 1 | 0 | 0 | 0 | 0 | 0 | 0 | 0 |
| ExuR/FucR | 1 | 1 | 0 | 0 | 0 | 0 | 0 | 0 | 0 | 1 | 0 | 0 | 0 | -1 | 0 | 0 | -1 | 0 | 0 | 0 |
| deletion-2 | 0 | 0 | 0 | 1 | 0 | 1 | 1 | 0 | 0 | 0 | 0 | 0 | 0 | 0 | 0 | 1 | 0 | 0 | 0 | 0 |
| Crp-2 | 0 | 0 | 1 | -1 | 0 | 0 | 0 | 0 | -1 | 0 | 0 | 0 | 1 | 0 | 0 | 0 | 0 | 0 | -1 | 0 |
| GlcC | 0 | 0 | 0 | 1 | 1 | 0 | 0 | 0 | 0 | 0 | -1 | -1 | 0 | 0 | 0 | 0 | 0 | 0 | 1 | 0 |
| insertion | 1 | 0 | 0 | -1 | 0 | 0 | 0 | 0 | 0 | 0 | 0 | 1 | 0 | 0 | 0 | 0 | 0 | 0 | 0 | 1 |
| FadR | 0 | 0 | 0 | 1 | 1 | 0 | 0 | 0 | 0 | 0 | 0 | 0 | 0 | 0 | 1 | 0 | 0 | 1 | 1 | 0 |
| Cra | 0 | 0 | 0 | 0 | -1 | 1 | 0 | 0 | 0 | 0 | 1 | 1 | 0 | 0 | 0 | 0 | 0 | 0 | -1 | 0 |
| purR-KO | 0 | 0 | 0 | 0 | -1 | 0 | 0 | 0 | 0 | 0 | 1 | 1 | 0 | 0 | 0 | 0 | 1 | 0 | 0 | 0 |
| Nac | 0 | 0 | 1 | 0 | 1 | 0 | 0 | 0 | 0 | 0 | 0 | 0 | 0 | 0 | 1 | 0 | -1 | 0 | 0 | 0 |
| NtrC+RpoN | 0 | 0 | 0 | 0 | 1 | 0 | 1 | 0 | 0 | 0 | -1 | -1 | 0 | 0 | 0 | 0 | 0 | 0 | 0 | 0 |
| GntR/TyrR | 0 | 1 | 0 | 0 | 0 | 0 | 1 | 0 | 0 | 1 | 0 | 0 | 0 | -1 | 0 | 0 | 0 | 0 | 0 | 0 |
| FecI | 0 | 0 | 0 | 0 | 0 | 1 | 0 | 0 | 0 | -1 | 0 | 0 | 0 | 0 | 0 | 1 | 1 | 0 | 0 | 0 |
| FliA | 0 | 0 | 0 | 0 | 0 | 0 | 1 | 0 | 0 | 0 | 0 | 0 | 0 | 1 | 0 | 0 | 0 | 0 | -1 | 0 |
| CpxR | 0 | 0 | 0 | 0 | 0 | -1 | -1 | 0 | 0 | 0 | 0 | 0 | 0 | -1 | 0 | -1 | 0 | 0 | 0 | 0 |
| uncharacterized-1 | 0 | 0 | 0 | 0 | 0 | 0 | 0 | 0 | 1 | 0 | 1 | 0 | 0 | 1 | 0 | 1 | 0 | 0 | 0 | 0 |
| MetJ | 0 | 0 | 0 | 0 | 0 | 0 | 0 | 0 | 0 | 0 | -1 | -1 | 0 | 0 | 0 | 0 | 1 | 1 | 0 | 0 |
| PrpR | 0 | 0 | 0 | 1 | 0 | 0 | 0 | 0 | 0 | 0 | 0 | -1 | 0 | 0 | 0 | -1 | 0 | 0 | 0 | -1 |
| GcvA | 0 | 0 | 0 | 0 | 0 | 0 | 0 | 0 | 0 | 0 | 0 | 0 | 0 | 0 | 0 | -1 | -1 | -1 | -1 | -1 |
| uncharacterized-4 | 0 | 0 | 0 | 0 | -1 | 0 | 0 | 0 | 0 | 0 | 1 | 1 | 0 | 0 | 0 | 0 | 0 | 0 | 0 | 0 |
| duplication-1 | 0 | 0 | 0 | 0 | 1 | -1 | -1 | 0 | 0 | 0 | 0 | 0 | 0 | 0 | 0 | 0 | 0 | 0 | 0 | 0 |
| YieP | 0 | 0 | 0 | 0 | -1 | 0 | 0 | 0 | 0 | 0 | 0 | 1 | 0 | 0 | 0 | 0 | 0 | 0 | -1 | 0 |
| RcsAB | 0 | 0 | 0 | 0 | -1 | 0 | 1 | 1 | 0 | 0 | 0 | 0 | 0 | 0 | 0 | 0 | 0 | 0 | 0 | 0 |
| thrA-KO | 0 | 0 | -1 | 0 | 0 | 0 | 0 | 0 | 0 | 0 | 0 | 0 | 0 | 0 | 1 | 0 | 0 | 0 | 0 | 1 |
| YgbI | 0 | 0 | 0 | 0 | 0 | 0 | 0 | 0 | 0 | 0 | 0 | 0 | 0 | 0 | 0 | 0 | 0 | -1 | -1 | -1 |
| uncharacterized-2 | 0 | 0 | 0 | 0 | 0 | 0 | 0 | 0 | 0 | 0 | 0 | 0 | 0 | 0 | 0 | 0 | 0 | 1 | 1 | 1 |
| GlpR | 0 | 1 | 0 | 0 | 0 | 0 | 0 | 0 | 0 | 0 | 0 | 0 | 0 | 0 | 0 | -1 | 0 | 0 | -1 | 0 |
| flu-yeeRS | 0 | 0 | 0 | 0 | 0 | 0 | 0 | 1 | 0 | 0 | 0 | 0 | 0 | 0 | 0 | 0 | 0 | 1 | 0 | 0 |
| ydcI-KO | 0 | 0 | 0 | 0 | 0 | 0 | 0 | 0 | 0 | 0 | 0 | 0 | 0 | 0 | 0 | 0 | 0 | 1 | 0 | 0 |
| Pyruvate | 0 | 1 | 0 | 0 | 1 | 0 | 0 | 0 | 0 | 0 | 0 | 0 | 0 | 0 | 0 | 0 | 0 | 0 | 0 | 0 |
| YiaJ | 1 | 1 | 0 | 0 | 0 | 0 | 0 | 0 | 0 | 0 | 0 | 0 | 0 | 0 | 0 | 0 | 0 | 0 | 0 | 0 |
| SrlR+GutM | 0 | 0 | 0 | 0 | 0 | 0 | 0 | 0 | -1 | 0 | -1 | 0 | 0 | 0 | 0 | 0 | 0 | 0 | 0 | 0 |
| Copper | 0 | 0 | 0 | 0 | 0 | 0 | 0 | 0 | -1 | 0 | 0 | 0 | 0 | 0 | 0 | -1 | 0 | 0 | 0 | 0 |
| XylR | 0 | 1 | 0 | 0 | 0 | 0 | 0 | 0 | 0 | 0 | 0 | 0 | 0 | -1 | 0 | 0 | 0 | 0 | 0 | 0 |
| lipopolysaccharide | 0 | 0 | 1 | 0 | 0 | 0 | 0 | 0 | 0 | 0 | 1 | 0 | 0 | 0 | 0 | 0 | 0 | 0 | 0 | 0 |
| entC-menF-KO | 0 | 0 | 0 | 1 | 0 | 0 | 0 | 0 | 1 | 0 | 0 | 0 | 0 | 0 | 0 | 0 | 0 | 0 | 0 | 0 |
| CsqR | 0 | 0 | 0 | 0 | 0 | 1 | 1 | 0 | 0 | 0 | 0 | 0 | 0 | 0 | 0 | 0 | 0 | 0 | 0 | 0 |
| nitrate-related | 0 | 0 | 0 | 0 | 0 | 0 | 0 | 0 | 0 | 0 | 0 | 0 | 0 | 0 | 0 | 0 | -1 | 0 | 0 | 0 |
| crp-KO | 0 | 0 | 0 | 0 | 0 | 0 | 0 | 0 | 0 | 0 | 0 | 0 | 0 | 0 | 0 | 0 | 1 | 0 | 0 | 0 |
| AllR/AraC/FucR | 0 | 1 | 0 | 0 | 0 | 0 | 0 | 0 | 0 | 0 | 0 | 0 | 0 | 0 | 0 | 0 | 0 | 0 | 0 | 0 |
| EvgA | 0 | 0 | 0 | 1 | 0 | 0 | 0 | 0 | 0 | 0 | 0 | 0 | 0 | 0 | 0 | 0 | 0 | 0 | 0 | 0 |
| YneJ | 0 | 0 | 0 | 1 | 0 | 0 | 0 | 0 | 0 | 0 | 0 | 0 | 0 | 0 | 0 | 0 | 0 | 0 | 0 | 0 |

**Table S9. Correlations between transcriptional responses to mutations.** Fold change in gene expression was assessed between each Mut strain and its Ref parent. The color-coded correlation coefficients between pairs of log2 fold changes are reported in the table. Strain names are color coded by their growth in M9G relative to the parent strain using the same scheme defined in Fig. 1. The information in this table is reported in Fig. 3.

|  |  |  | Rif | | | | | | | | | | M9G | | | | | | | | | |
|---|---|---|---|---|---|---|---|---|---|---|---|---|---|---|---|---|---|---|---|---|---|---|
|  |  |  | pykF(C8Y) | | | | | rpoB(T1037P) | | | | | pykF(C8Y) | | | | | rpoB(T1037P) | | | | |
|  |  |  | Q148L | L511R | I572S | D516G | H526Y | Q148L | I572F | L511R | I572N | S508P | Q148L | L511R | I572S | D516G | H526Y | Q148L | I572F | L511R | I572N | S508P |
| Rif | pykF(C8Y) | Q148L | 1.00 | 0.49 | 0.19 | -0.23 | -0.32 | 0.39 | 0.21 | 0.27 | 0.00 | 0.05 | 0.34 | 0.24 | -0.17 | -0.37 | -0.40 | 0.19 | 0.06 | 0.06 | -0.22 | -0.23 |
| | | L511R | 0.49 | 1.00 | 0.17 | -0.12 | -0.30 | 0.32 | 0.37 | 0.39 | 0.19 | 0.07 | 0.22 | 0.59 | -0.13 | -0.45 | -0.61 | 0.09 | 0.21 | 0.20 | -0.14 | -0.24 |
| | | I572S | 0.19 | 0.17 | 1.00 | 0.12 | 0.06 | 0.08 | 0.13 | 0.15 | 0.22 | 0.15 | -0.01 | 0.02 | 0.11 | 0.02 | -0.04 | -0.01 | 0.00 | 0.01 | 0.03 | 0.00 |
| | | D516G | -0.23 | -0.12 | 0.12 | 1.00 | 0.79 | -0.50 | 0.01 | -0.27 | 0.29 | -0.05 | -0.27 | -0.16 | 0.70 | 0.68 | 0.44 | -0.14 | 0.07 | 0.03 | 0.31 | 0.39 |
| | | H526Y | -0.32 | -0.30 | 0.06 | 0.79 | 1.00 | -0.47 | -0.10 | -0.32 | 0.25 | 0.11 | -0.29 | -0.37 | 0.43 | 0.69 | 0.67 | -0.19 | -0.06 | -0.12 | 0.28 | 0.51 |
| | rpoB(T1037P) | Q148L | 0.39 | 0.32 | 0.08 | -0.50 | -0.47 | 1.00 | 0.60 | 0.82 | 0.36 | 0.54 | 0.25 | 0.11 | -0.53 | -0.49 | -0.44 | 0.24 | 0.04 | 0.17 | -0.13 | -0.34 |
| | | I572F | 0.21 | 0.37 | 0.13 | 0.01 | -0.10 | 0.60 | 1.00 | 0.80 | 0.77 | 0.54 | 0.08 | 0.19 | -0.07 | -0.19 | -0.37 | 0.07 | 0.49 | 0.39 | 0.25 | -0.13 |
| | | L511R | 0.27 | 0.39 | 0.15 | -0.27 | -0.32 | 0.82 | 0.80 | 1.00 | 0.56 | 0.63 | 0.15 | 0.23 | -0.37 | -0.31 | -0.39 | 0.13 | 0.26 | 0.42 | 0.14 | -0.19 |
| | | I572N | 0.00 | 0.19 | 0.22 | 0.29 | 0.25 | 0.36 | 0.77 | 0.56 | 1.00 | 0.53 | -0.08 | 0.02 | 0.16 | 0.10 | -0.07 | -0.01 | 0.45 | 0.23 | 0.47 | 0.11 |
| | | S508P | 0.05 | 0.07 | 0.15 | -0.05 | 0.11 | 0.54 | 0.54 | 0.63 | 0.53 | 1.00 | 0.05 | -0.09 | -0.29 | -0.01 | 0.02 | -0.01 | 0.08 | 0.14 | 0.18 | 0.18 |
| M9G | pykF(C8Y) | Q148L | 0.34 | 0.22 | -0.01 | -0.27 | -0.29 | 0.25 | 0.08 | 0.15 | -0.08 | 0.05 | 1.00 | 0.20 | -0.19 | -0.24 | -0.27 | 0.15 | 0.03 | 0.09 | -0.14 | -0.17 |
| | | L511R | 0.24 | 0.59 | 0.02 | -0.16 | -0.37 | 0.11 | 0.19 | 0.23 | 0.02 | -0.09 | 0.20 | 1.00 | 0.02 | -0.25 | -0.43 | 0.13 | 0.38 | 0.27 | 0.13 | -0.09 |
| | | I572S | -0.17 | -0.13 | 0.11 | 0.70 | 0.43 | -0.53 | -0.07 | -0.37 | 0.16 | -0.29 | -0.19 | 0.02 | 1.00 | 0.50 | 0.31 | -0.07 | 0.19 | 0.04 | 0.26 | 0.21 |
| | | D516G | -0.37 | -0.45 | 0.02 | 0.68 | 0.69 | -0.49 | -0.19 | -0.31 | 0.10 | -0.01 | -0.24 | -0.25 | 0.50 | 1.00 | 0.75 | -0.15 | 0.09 | -0.12 | 0.47 | 0.51 |
| | | H526Y | -0.40 | -0.61 | -0.04 | 0.44 | 0.67 | -0.44 | -0.37 | -0.39 | -0.07 | 0.02 | -0.27 | -0.43 | 0.31 | 0.75 | 1.00 | -0.14 | -0.22 | -0.28 | 0.27 | 0.53 |
| | rpoB(T1037P) | Q148L | 0.19 | 0.09 | -0.01 | -0.14 | -0.19 | 0.24 | 0.07 | 0.13 | -0.01 | -0.01 | 0.15 | 0.13 | -0.07 | -0.15 | -0.14 | 1.00 | 0.05 | 0.09 | 0.01 | -0.10 |
| | | I572F | 0.06 | 0.21 | 0.00 | 0.07 | -0.06 | 0.04 | 0.49 | 0.26 | 0.45 | 0.08 | 0.03 | 0.38 | 0.19 | 0.09 | -0.22 | 0.05 | 1.00 | 0.33 | 0.68 | 0.12 |
| | | L511R | 0.06 | 0.20 | 0.01 | 0.03 | -0.12 | 0.17 | 0.39 | 0.42 | 0.23 | 0.14 | 0.09 | 0.27 | 0.04 | -0.12 | -0.28 | 0.09 | 0.33 | 1.00 | 0.24 | -0.01 |
| | | I572N | -0.22 | -0.14 | 0.03 | 0.31 | 0.28 | -0.13 | 0.25 | 0.14 | 0.47 | 0.18 | -0.14 | 0.13 | 0.26 | 0.47 | 0.27 | 0.01 | 0.68 | 0.24 | 1.00 | 0.45 |
| | | S508P | -0.23 | -0.24 | 0.00 | 0.39 | 0.51 | -0.34 | -0.13 | -0.19 | 0.11 | 0.18 | -0.17 | -0.09 | 0.21 | 0.51 | 0.53 | -0.10 | 0.12 | -0.01 | 0.45 | 1.00 |